\documentclass{aa}
\def\b{\bigskip}

\def\ki2{$\chi^2$}
\def\lya{Lyman $\alpha$}
\def\kms{${\rm km.s}^{-1}$}
\def\cm2{${\rm cm}^{-2}$}
\def\dsh{D/H}
\input{psfig}

\def\la{\mathrel{\mathchoice {\vcenter{\offinterlineskip\halign{\hfil
$\displaystyle##$\hfil\cr<\cr\sim\cr}}}
{\vcenter{\offinterlineskip\halign{\hfil$\textstyle##$\hfil\cr<\cr\sim\cr}}}
{\vcenter{\offinterlineskip\halign{\hfil$\scriptstyle##$\hfil\cr<\cr\sim\cr}}}
{\vcenter{\offinterlineskip\halign{\hfil$\scriptscriptstyle##$\hfil
\cr<\cr\sim\cr}}}}}

\begin{document}

\newcommand{\excs}{\extracolsep{\fill}}
\newcommand{\msun}{M_\odot}
\newcommand{\eqr}[1]{Eq.~(\ref{#1})}
\newcommand{\dy}{\displaystyle}
\newcommand{\gbb}{G191-B2B}
\newcommand{\Cap}{Capella}
\newcommand{\h}{\mbox{H}}
\newcommand{\hi}{\mbox{H\,{\sc i}}}
\newcommand{\hii}{\mbox{H\,{\sc ii}}}
\newcommand{\di}{\mbox{D\,{\sc i}}}
\newcommand{\oi}{\mbox{O\,{\sc i}}}
\newcommand{\Ni}{\mbox{N\,{\sc i}}}
\newcommand{\sid}{\mbox{Si\,{\sc ii}}}
\newcommand{\sit}{\mbox{Si\,{\sc iii}}}

\thesaurus{08.09.2; 09.01.1; 13.21.3}
\titlerunning{Detection of local ISM D/H variations.}
\authorrunning{Vidal-Madjar {\it et al.}}
\title{Detection of spatial variations in the (D/H) ratio in the
local interstellar medium.
\thanks{Based on observations with the NASA/ESA Hubble Space Telescope, obtained
at the Hubble Space Telescope Science Institute which is operated by the
Association of Universities for Research in Astronomy Inc., under NASA contract
NAS5-26555.}
}

\author{Alfred Vidal-Madjar\inst{1}
\and Martin Lemoine\inst{2}
\and Roger Ferlet\inst{1}
\and Guillaume~H\'ebrard\inst{1}
\and Detlev Koester\inst{3}
\and Jean Audouze\inst{1}
\and Michel Cass\'e\inst{4,1}
\and Elisabeth Vangioni-Flam\inst{1}
\and John~Webb\inst{5}
}
\institute{Institut d'Astrophysique de Paris, CNRS, 98 bis boulevard Arago,
75014 Paris, France
	\and DARC, UPR-176 CNRS, Observatoire de Paris-Meudon, 92195 Meudon
C\'edex, France
	\and Universitaet Kiel, Germany
        \and CEA/DSM/DAPNIA, Service d'Astrophysique, Saclay, 91191
Gif-sur-Yvette, France
	\and School of Physics, University of New South Wales, Sydney, NSW
2052, Australia
}
\date{Received ....; accepted ....}
\offprints{A. Vidal-Madjar}
\maketitle
\begin{abstract}
  We present high resolution ($\Delta\lambda\simeq3.7$\kms)
HST-GHRS observations
of the DA white dwarf G191-B2B, and derive the interstellar \dsh\ ratio on the
line of sight. We have observed and analysed simultaneously the
interstellar lines of \hi, \di, \Ni, \oi, \sid\ and \sit.
We detect three absorbing clouds, and derive a
total \hi\ column density N(\hi)=2.4$\pm0.1 \times10^{18}$cm$^{-2}$,
confirming our Cycle~1 estimate, but in disagreement with other
previous measurements.

We derive an average \dsh\ ratio over the three absorbing clouds
N(\di)$_{\rm total}$/N(\hi)$_{\rm total}$=1.12$\pm 0.08~\times 10^{-5}$,
in  disagreement with the previously reported value of the local
\dsh\ as reported by Linsky {\it et al.} (1995) toward Capella.
We re-analyze the GHRS data of the Capella line of sight, and
confirm their estimate, as we find 
(\dsh)$_{\rm Capella}=1.56\pm 0.1~\times 10^{-5}$ in
the Local Interstellar Cloud in which the solar system is embedded. 
This shows that the \dsh\ ratio varies by at least $\sim30\%$ within the local
interstellar medium.

Furthermore, the Local Interstellar Cloud is also detected toward \gbb,
and we show that the \dsh\ ratio in this component, toward \gbb, can be made
compatible with that derived toward Capella.
However, this comes at the expense of a much smaller
value for the \dsh\ ratio as averaged over the other two components, of order
$0.9\times10^{-5}$, and in such a way that the \dsh\ ratio as averaged over
all three components remains at the above value, {\it i.e.}
(\dsh)$_{\rm Total}=1.12\times10^{-5}$.

We thus conclude that, either the \dsh\ ratio varies from cloud to cloud, 
and/or the \dsh\ ratio varies within the Local Interstellar Cloud, in which
the Sun is embedded, although our observations
neither prove nor disprove this latter possibility.

\end{abstract}

\section{Introduction}

Deuterium is only produced in primordial Big Bang nucleosynthesis (BBN),
and  destroyed in stellar interiors (Epstein, Lattimer \&
Schramm 1976). Hence, any abundance of deuterium measured at any
metallicity should provide a lower limit to the primordial deuterium
abundance (Reeves {\it et al.} 1972). Deuterium is thus a key element
in cosmology and in galactic
chemical evolution ({\it e.g.}, Steigman, Schramm \& Gunn 1977; Audouze
\& Tinsley 1976; Vidal-Madjar \& Gry 1984; Boesgaard \& Steigman 1985; 
Olive {\it et al.} 1990; Vangioni-Flam \& Cass\'e 1995; Prantzos 1996, 
Scully {\it et al.} 1997).
The primordial abundance of deuterium is indeed the best probe of the
baryonic density parameter of the Universe $\Omega_B$, and the decrease of
its abundance with  galactic evolution traces, amongst other things,
the amount of star formation.

The first, although indirect, measurement of the deuterium abundance of 
astrophysical significance was carried out through $^3$He  in
the solar wind, leading to \dsh$\simeq2.5\pm1.0\times10^{-5}$ (Geiss \&
Reeves 1972), a value representative of an epoch 4.5 Gyrs past. The first
measurements of the interstellar \dsh\ ratio, representative of the present
epoch, were reported shortly thereafter (Rogerson \& York 1973).
Their value of \dsh$\simeq1.4\pm0.2\times10^{-5}$ has not changed ever since.
The most accurate measurement of the interstellar \dsh\ ratio was reported
by Linsky {\it et al.} (1993, 1995, hereafter L93, L95) 
in the direction of Capella, using HST-GHRS,
\dsh$\simeq1.6\pm0.1\times10^{-5}$ (statistical + sytematic). 

Up to a few years ago, these measurements were used to constrain BBN in a
direct way. The situation has changed, as measurements of the \dsh\ ratio
in metal-deficient quasars absorbers, at moderate and high redshift, have
become available ({\it e.g.}, Carswell {\it et al.} 1994; Songaila
{\it et al.} 1994; Tytler, Fan \& Burles 1996; 
Webb {\it et al.} 1997; Burles \& Tytler 1998a,b; see also Burles \& Tytler,
1998c for a review). However, these
observations have not provided a single definite value of the primordial
\dsh\ ratio. At the present time, it is not known whether the higher
estimates of the primordial \dsh\ ratio reported are artifacts due to the
mimicking of the \di\ line by an \hi\ interloper, or whether the ratios
reported (low or high) are discrepent due to errors in interpreting the
velocity structure, or whether substantial fluctuations
(a factor $\sim6$) actually exist. 

On similar grounds, it turns out that determinations of the interstellar \dsh\
ratio do not generally agree on a single value, even in the very local
medium (Vidal-Madjar {\it et al.}  1986, Murthy {\it et al.}  1987,
1990). For instance, \dsh$<10^{-5}$ is measured toward $\lambda$ Sco (York,
1983), while in the opposite direction toward $\alpha$ Aur,
\dsh=$1.65\times10^{-5}$ (L93, L95). On longer pathlengths, 
\dsh$\simeq7.\times10^{-6}$ is measured toward $\delta$ and $\epsilon$ Ori
(Laurent {\it et al.}  1979), and 
\dsh$\simeq5.\times10^{-6}$ toward $\theta$ Car
(Allen {\it et al.}  1992). Although several scenarios have been proposed
to explain these putative variations ({\it e.g.},
Vidal-Madjar {\it et al.}  1978; Bruston {\it et al.}  1981), the above 
measurements are still unaccounted for. Despite the high quality of the data,
it may nevertheless be possible that even better data is required to overcome
systematic effects.

The only way to derive a reliable estimate of the interstellar \dsh\ 
ratio is to observe the atomic transitions of D and H in 
the far-UV, in absorption in the local ISM against the background continuum
of cool or hot stars. These observations have been performed using the
Copernicus and IUE satellites, and now the Hubble Space Telescope. Both types
of target stars present pros and cons.

The main advantage of observing cool stars is that they can be selected in
the vicinity of the Sun. This results in low H{\sc i} column densities, and 
simple lines of sight. A difficulty inherent to the
cool stars approach is that the detailed structure 
of the line of sight can be obtained only through the observation of the 
Fe{\sc ii} and the Mg{\sc ii} ions, which are unfortunately not proper
tracers of \hi. In particular, species like N{\sc i} and O{\sc i} could not
be observed. Moreover, the estimate of the \hi\ column density
always depends strongly on the modeling of the chromospheric Lyman $\alpha$
emission line. The Capella target of L93, L95 is a
double system with two cool stars: 
it provided a line of sight with a single absorbing component (see
however Sec.5), and allowed a very accurate estimate of the \dsh\ ratio, as
the emission line could be modeled by observing the binary system at
different phases.

Hot stars are unfortunately located further away from the Sun, so that one 
always has to face a high H{\sc i} column density and often a non-trivial 
line of sight structure. In these cases, D{\sc i} could not be detected at 
Ly$\alpha$, and one has to observe higher order lines, {\it e.g.} Ly$\gamma$,
Ly$\delta$, Ly$\epsilon$. The stellar continuum is however smooth at the 
location of the interstellar absorption and, moreover, the N{\sc i} triplet 
at 1200~\AA ~as well as other N{\sc i} lines are available to probe the 
velocity structure of the line of sight.  N{\sc i} and O{\sc i} 
were shown to be reliable tracers of H{\sc i} in the ISM (Ferlet 1981; 
York {\it et al.} 1983). 

We introduced in Cycle 1 of HST a new type of target,  white dwarfs,
which should solve many of the intrinsic difficulties of the problem.
Indeed, such targets may be chosen in the high
temperature range where the depth of the photospheric Lyman $\alpha$ line is
reduced, so as to obtain a significant flux at the bottom of the \hi\ stellar
absorption line where the interstellar \di\ and \hi\ lines appear, as well as
a smooth stellar continuum. Also, these targets may be chosen
close to the Sun so that the line of sight is not too complex. We observed in
HST Cycle 1 the white dwarf \gbb, at medium resolution
$\Delta\lambda\simeq18$\kms\ (Lemoine {\it et al.} 1996).

We proceeded further in this approach with Cycle 5 high resolution
($\Delta\lambda\simeq3.7$\kms) Echelle-A data  of \hi, \di, as well as
\Ni\ and \oi, and \sid\ and \sit, in order to derive an accurate velocity
and ionization structure of the line of sight.
We present these new spectroscopic observations of the white dwarf 
G191--B2B in Section 2. In Section 3 we present the
analysis of these data; the results are further analysed in Section 4. We
re-analyze the Capella line of sight in Section 5, and summarize our
conclusions in Section 6.
\b

\section{Observations and Data Reduction.}
\subsection{Observations}

\begin{table*}[tp]
\caption[]{List of lines observed in 1995 during Cycle 5 at high
spectral resolution $\Delta\lambda\simeq3.5$\kms.}
\label{line}
\begin{tabular*}{\textwidth}
{l@{\excs}l@{\excs}l@{\excs}l@{\excs}l@{\excs}l@{\excs}l@{\excs}l}
\hline
File & Element & Central & Observing & Grating & Exp. & S/N\\
& &  Wave.  & Date & & Time  & \\
& &   (\AA) &  (1995) &  & (s)  &\\
\hline
%z2q3010ft & \hi, \di, \Ni, \sid, \sit &  1215.71 & Jul 26, 20:00 &  G140M & 870  & 10\\
%z2q3010gt & \hi, \di, \Ni, \sid, \sit &  1199.21 & Jul 26, 20:18 &  G140M & 1958 & 10\\
%z2q3010ht & \hi, \di, \Ni, \sid, \sit &  1215.71 & Jul 26, 21:14 &  G140M & 544  & 10\\
z2q30208m & \hi, \di                  &  1215.45 & Jul 26, 22:18 &  ECH-A & 6528 & 21\\
z2q3020ct & \hi, \di                  &  1215.45 & Jul 27, 00:59 &  ECH-A & 6528 & 21\\
z2q3020gt & \hi, \di                  &  1215.45 & Jul 27, 03:39 &  ECH-A & 6528 & 21\\
z2q30108t & \Ni                       &  1199.94 & Jul 26, 13:30 &  ECH-A & 4896 & 25\\
z2q3010dt & \Ni, \sit                 &  1203.57 & Jul 26, 18:27 &  ECH-A & 3917 & 19\\
z2q3010cm & \oi, \sid                 &  1303.01 & Jul 26, 16:31 &  ECH-A & 4570 & 26\\
\hline
\end{tabular*}\\
\end{table*}

Our new observations of the white dwarf G191--B2B were performed with
GHRS/HST in July 1995 (Cycle 5 Guest Observer proposal ID5893).
The spectra were acquired in the wavelength ranges 
1196--1203\AA,
1200--1207\AA,
1212--1219\AA\ and
1299--1307\AA\ using the Echelle-A grating.
The log of these observations is presented in Table~\ref{line}.

The Echelle-A grating provides a nominal resolving power of $\sim$80~000,
or a spectral resolution of 3.7 \kms . 
We used only the Small Science Aperture (SSA), corresponding to
0.25" on the sky and illuminating 1 diode to achieve the best possible
resolving power. For further details on the
instrumentation, see Duncan (1992).

G191--B2B is a DA spectral type white dwarf located 40-70 pc away from the
Sun, with an effective temperature T$_{\rm eff}=61190-61700$K, gravity 
Log(g)=$7.49-7.61$ (Finley {\it et al.} 1997; Vauclair {\it et al.} 1997),
and magnitude m$_v$=11.8, as determined from optical data fitted with pure
hydrogen models. It fulfills all the criteria for a
good candidate to measure the \dsh\ ratio. Apart from \di\ and \hi\ at
Lyman $\alpha$, we obtained the spectra of the lines
\Ni(1199\AA, 1200\AA, 1201\AA), 
\sit (1206\AA), \oi (1302\AA) and 
\sid (1304\AA) using Echelle-A. The observation at high spectral
resolution of \oi\ and \Ni\ should
allow us to resolve possible different \hi\ absorbing clouds on the 
line of sight, while the observation of \sid\ and \sit\ should allow the
detection of possible \hii\ gas. From the comparison of the $b$~values
of different atomic weights, an estimate of the
temperature and turbulent velocity in the different clouds is also
possible, while the analysis of the \sid\ and \sit\ lines should
determine the ionization structure of the line of sight. In particular, the
observed lines of \Ni, \sid\ and \sit\ are not saturated, and should therefore
provide accurate estimates of the column densities. The \oi\ 1302\AA\ line
is slightly saturated in this low \hi\ column density environment, and should
thus offer a reasonable estimate of the column densities, as well as of the
$b$-values. At the end, we expect to obtain a sufficient number of constraints
on all parameters involved in the analysis of the Lyman $\alpha$ lines of \di\ and
\hi, but the \dsh\ ratios of the different components.

\subsection{Data reduction.}

Our data were reduced with the Image Reduction and Analysis Facility (IRAF)
software, using the STSDAS package.
During the observations, we used the FP-SPLIT mode which splits the total
exposure time into successive cycles of 4 sub-exposures, each corresponding to
a slightly different projection of the spectrum on the photocathode. We used
the ``quarter stepping'' mode, which provides a sample of 4 pixels per
resolution element. This oversamples
the spectrum (since for instance the SSA does not fulfill the 
Nyquist sampling criterion), and allows to correct for the granularity of the
photocathode. Indeed, the effect of the photocathode on each diode being 
the same for the four sub-exposures, it is possible to evaluate this granularity 
from the comparison of the 4 sub-exposures where a constant granularity effect
mixes with
a non-constant photon statistical noise. 

We found that the standard method for
correcting for the granularity, which is available in the IRAF-STSDAS package,
was not efficient as a result of the impact of the geomagnetic field on the
photocathode. The geomagnetic field deflects the incident electron beam
on the diodes
in a way depending on the location of the HST on its orbit, thus making the
granularity pattern non-constant from sub-exposure to sub-exposure. As a 
result,
the standard iterative method leads to the appearance of noisy features whose
intensity increases with the number of iterations. We therefore developed a
different procedure based on the use of simple statistical filters.

\begin{figure}[h]
\setlength{\unitlength}{1cm}
\centering
%\begin{picture}(9,9)
%\put(0,0){\makebox(8,10){
\psfig{figure=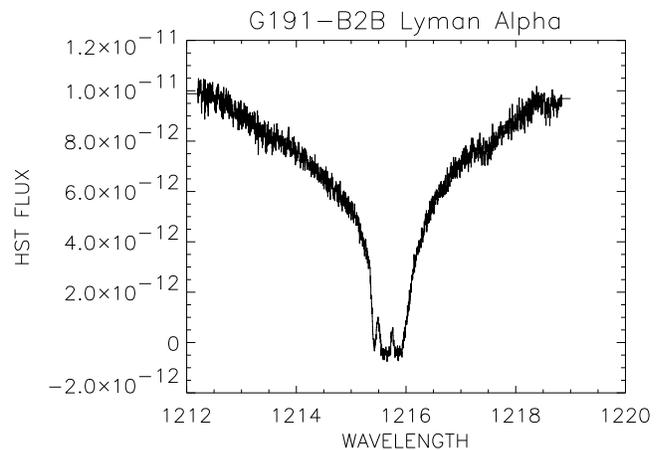,width=\columnwidth}
%}}
%\end{picture}
\caption{Data taken with the Echelle-A in the \lya\ region.
Some perturbations due to the photocathode inhomogeneities can be seen
near 1213.4 and 1217.5 \AA. Note the deuterium absorption at 1215.3\AA. 
The central peak at the bottom of the saturated \hi\ \lya\ line
corresponds to the earth geocoronal emission. Note also that the bottom of
the saturated \lya\ line is not at exact zero level.}
\label{HI}
\end{figure}

We first cross-correlate the different spectra corresponding to the same spectral
ranges, {\it i.e.} the different sub-exposures of a same exposure, and shift
them accordingly so that real absorption or emission features correspond to the
same pixel values in each spectrum. We next compute the median and the standard
deviation with respect to that median at each pixel, our statistical sample
being the set of  intensities of the different spectra at that pixel.
Finally, we compute the average intensity value at each pixel, rejecting the
intensities of the different spectra deviating from the median by more than,
say, 3 standard deviations. We found that this simple treatment resulted
in good signal-to-noise ratios (Fig.~\ref{HI}) while
avoiding any residual ghost feature (see Bertin {\it et al.} 1995 for more
details). 

We chose to rebin our spectra to two pixels per diode in all cases but \oi, 
{\it i.e.} two pixels per
resolution element approximately, which is the standard sampling mode, using
the IRAF-STSDAS procedure. In the case of \oi\, the very sharp edges of the
line  contain most of the information, and we thus decided to 
keep four pixels per resolution element so as to take advantage
of the oversampling available in these observations.

In the case of Lyman $\alpha$, the spectral signature is not as sharp, and,
moreover, the S/N ratio is significantly lower at the bottom of the line, and
we thus developed a second approach to check the consistency of our procedure.
We re-align the sub-exposures using the deuterium line, which is clearly
detected
in each of the sub-exposure, as a reference. This procedure means we are unable
to correct for the photocathode granularity. Nonetheless,
it appears that no such defect is
present at \lya. In effect, such defects are of order $\sim$10\%,
and are thus easily detected when adding the sub-exposures corresponding to
{\it a priori} the same spectral instrument shift,
thereby improving the S/N ratio and building four different shifted spectra
where the photocathode defects appear at fixed positions. This process
allowed us to show that the amplitude of photocathode defects in the \lya\
central region does not exceed a few percent. One clearly sees such defects in
Fig.~\ref{HI} in the far wings of the \lya\ profile, around 1213.4\AA\ and
1217.5\AA\ (these defects were not perfectly corrected through the first
approach). Thus the final \lya\ line profile is well
represented in the region from 1213.6\AA\ to 1217.3\AA, and the region that we
 use in our data analysis is even further reduced to
1214.5-1217.0\AA, where the  spectral information lies.

As a second consistency check, we measure the width of the geocoronal emission
line. Spurious shifts in the co-adding procedure would degrade the 
spectral
resolution. The geocoronal emission line should be a very narrow 
feature,
only slightly broader than the nominal instrument spectral resolution.
We find,  fitting the emission line with a gaussian, an equivalent Doppler
width $b\simeq5.7$\kms, corresponding to a temperature $T\simeq1950$K, after
deconvolution from the instrumental profile. At this point, one should note
that the geocorona fills up the SSA, so that its width results from the
convolution of
the intrinsic width with the spectrograph instrumental profile, and with the
profile of a point source broadened by the projection of the SSA on the
spectral
scale; this latter effect amounts to broadening the SSA instrumental profile by
4 pixels. The above measured width thus corresponds to an intrinsic 
temperature $T\sim600$K, which is a very plausible average value for the Earth
exosphere temperature. We note that this consistency test is all the more
significant as the geocoronal emission line only shows up in the final co-added
sub-exposures. Therefore, the procedure used for  co-addition  produces
a sharp line profile in which the Echelle-A spectral resolution is kept at its
highest nominal value.

\begin{figure}[h]
\setlength{\unitlength}{1cm}
\centering
%\begin{picture}(9,9)
%\put(0,0){\makebox(8,10){
\psfig{figure=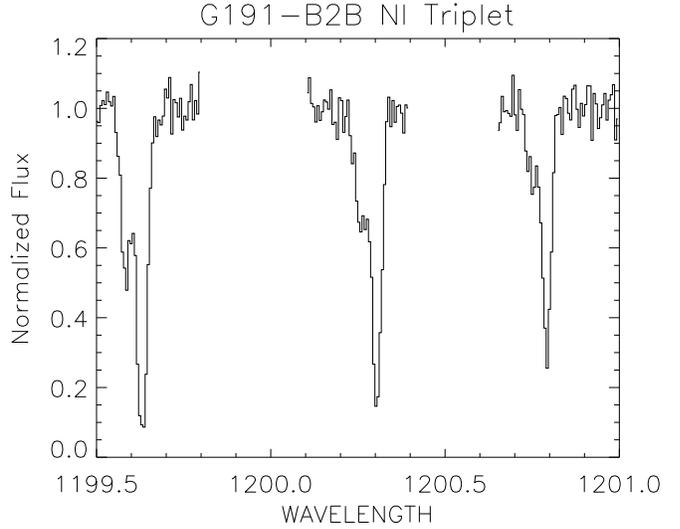,width=\columnwidth}
%}}
%\end{picture}
\caption{The normalized \Ni\ triplet data used in the fitting procedure.}
\label{NI}
\end{figure}
\begin{figure}[h]
\setlength{\unitlength}{1cm}
\centering
%\begin{picture}(9,9)
%\put(0,0){\makebox(8,10){
\psfig{figure=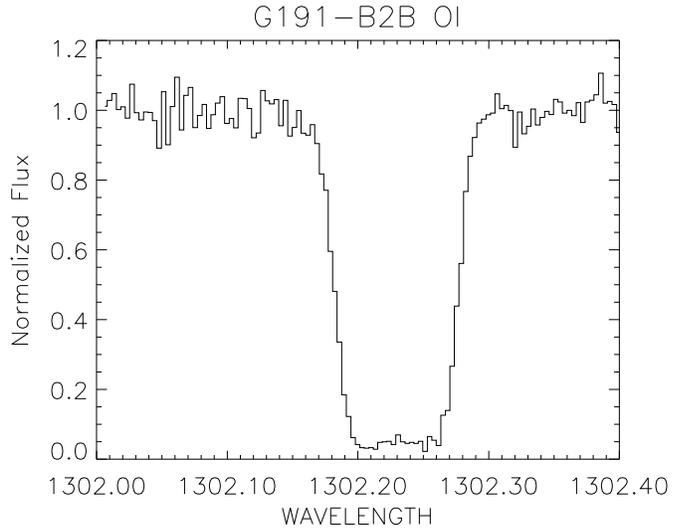,width=\columnwidth}
%}}
%\end{picture}
\caption{Same as Fig.~\ref{NI}. The \oi\ line.}
\label{OI}
\end{figure}
\begin{figure}[h]
\setlength{\unitlength}{1cm}
\centering
%\begin{picture}(9,9)
%\put(0,0){\makebox(8,10){
\psfig{figure=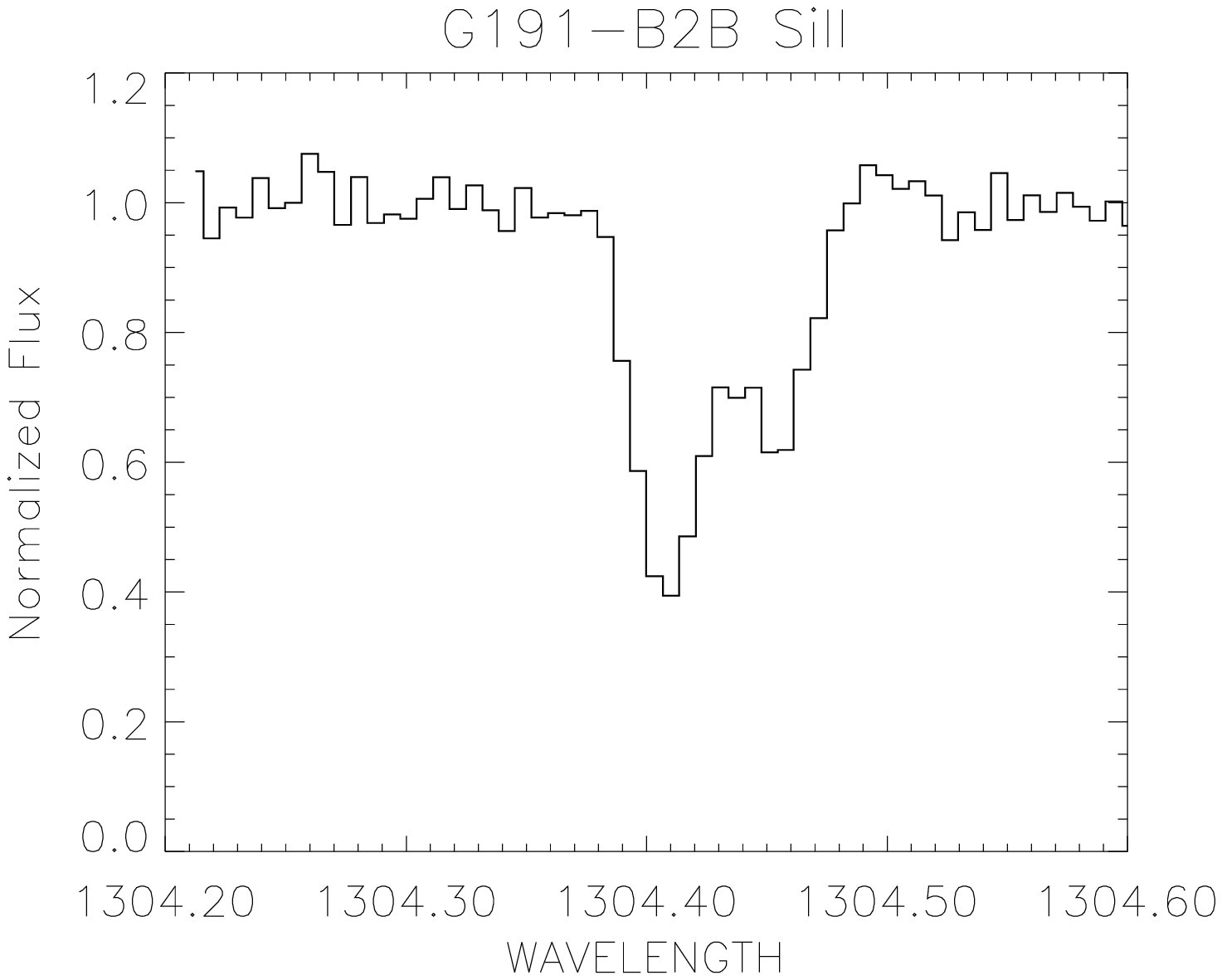,width=\columnwidth}
%}}
%\end{picture}
\caption{Same as Fig.~\ref{NI}. The \sid\ line.}
\label{SiII}
\end{figure}
\begin{figure}[h]
\setlength{\unitlength}{1cm}
\centering
%\begin{picture}(9,9)
%\put(0,0){\makebox(8,10){
\psfig{figure=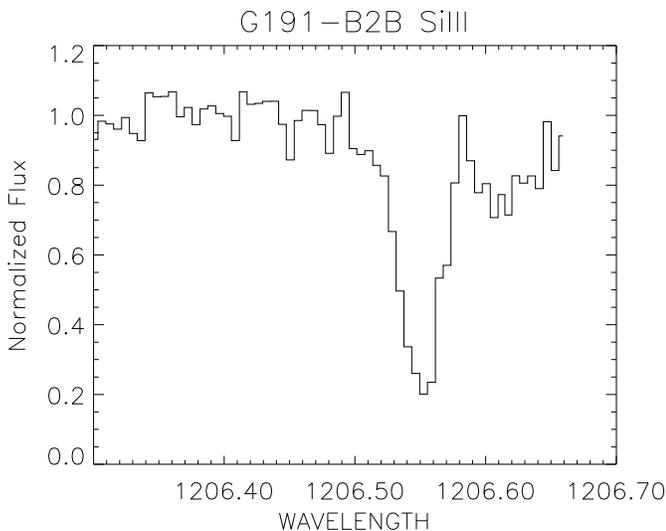,width=\columnwidth}
%}}
%\end{picture}
\caption{Same as Fig.~\ref{NI}. 
The \sit\ line. Note that the broad line on the red side is of photospheric 
origin (see text).}
\label{SiIII}
\end{figure}

Each exposure was preceded by a special platinum lamp calibration
which allowed us to calculate the
residual shift left by the standard CAL-HRS calibration procedure. 
Together with the oversampling mode, this correction
allowed us to reach an absolute calibration accuracy of $\pm$1.5\kms\ in 
radial velocity. Hereafter, the radial velocities
will be given in the heliocentric rest frame.
 
The different spectral lines were normalized to a continuum of unity by a
polynomial whose degree, typically $2-7$, was chosen according to a
cross-validation statistical procedure.  The normalized
spectra are shown
in Fig.~\ref{NI}, Fig.~\ref{OI}, Fig.~\ref{SiII} and Fig.~\ref{SiIII}. Since
the continuum level is determined to high precision, we do not need to take 
into account any continuum fitting parameters in the line fitting.
The  \lya\ line was not normalized in this way,
as the stellar continuum is a Lorentzian profile, with a deep core;
the stellar continuum at \lya\ is described at length later.

Finally, in the case of the \sit\ line (1206\AA), we decided 
to keep in the spectrum the nearby signature of the photospheric \sit\ 
stellar line on the red wing of the interstellar \sit\ line (see Fig.~\ref{SiIII}). 
This photospheric line was thus fitted along with the interstellar features as an
additional component. This gave access to an independent evaluation of the
intrinsic shift of the white dwarf photospheric features, a parameter that is also
involved in our final analysis of the deuterium and hydrogen \lya\ profiles.

\subsection{The number of absorption components along the line of sight}

\begin{figure}[t]
\psfig{file=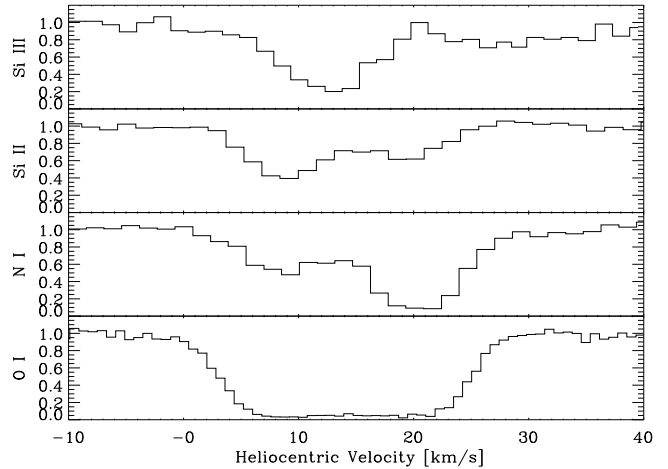,width=\columnwidth}
\caption{The absorption features seen in different species,
as indicated on the left hand side of the figure, superposed in
velocity space. Two obvious 
components are seen in \sid\ and \Ni. A third component in 
between the other two is detected in the \sit\ and \oi\ lines. The 
two components at the shortest wavelengths, seen in \sit, are \hii\ 
regions, while the component at $\sim$20 km s$^{-1}$ is the Local Cloud, an 
\hi\ region. It is obvious that deducing velocity structure of lines of 
sight for \dsh\ evaluations from ionized species could be extremely misleading.}
\label{components}
\end{figure}

We first evaluate the number of components present on the line of sight. 
This number represents a minimum number of components since blends
are always possible.

Figure~\ref{components} illustrates all species (excluding \hi\ and \di)
observed with the Echelle-A spectrograph. 
It is clear, from the \Ni\ and \sid\ profiles, that at least two
clearly resolved components are present. 
In the \sit\ line, the strongest component falls in between the
other two seen in \Ni\ and \sid.
This cannot be due to an incorrect absolute calibration of one of the spectral
ranges, as we observed the \sit\ line at 1206\AA\ together with the
\Ni\ line at 1201\AA\ in the same exposure. The relative shift between
these two species, from one end of the detector to the other,
should certainly be less than
0.5~\kms. This \sit\ component is thus real, and obviously more ionized than
the other two.

A broad component is also detected in \sit\ at the edge of the
observed spectral range. When fitted with a
single absorption feature, assuming that the broadening is thermal, the
corresponding temperature is of order 30000~K, {\it i.e.} much broader than
clouds usually seen in the local ISM. Because its heliocentric
velocity is $\simeq29$\kms, it is  reasonable to assume that this line
is of photospheric origin. Indeed, as shown in Vidal-Madjar {\it et al.}
(1994),  photospheric features in \gbb\ are relatively broad and their
observed heliocentric velocities are $\sim35\pm4$\kms . 
Predictions of the strength of the \sit\ photospheric line are also
in agreement with the observations. We thus do not include this feature in
our study.

The \oi\ profile confirms, through profile fitting, the need for a third
insterstellar component, located in between the two components seen in \Ni\ and
\sid. The \oi\ signature of this third component is mainly the flatness
of the absorption in the saturated core. We also note that there is no
component around 29 \kms in the other lines, consistent with the photospheric
interpretation of the broad \sit\ feature.

\subsection{The residual flux at the bottom of the spectral lines.}

In order to properly evaluate the strength of the different ISM
components contributing to the observed absorption, it is extremely
important to evaluate the level of the zero flux which may be slightly
erroneous in an echelle spectrograph, due to the diffuse light produced from
the adjacent spectral orders. The case of \lya\ is obvious as the broad
saturated core allows us to accurately determine for the zero level.
We corrected the \lya\ region for this zero level by subtracting the
flux in the \lya\ core from the whole line profile.
Correcting for the zero level in the GHRS data reduction pipeline is indeed
difficult for other species, as it depends on instrumental parameters
as well as on the (unknown)
spectrum shape in adjacent orders.  Thus, for species other than \hi,
we determined the zero level as the best
fit value, {\it i.e.} the one corresponding to a minimum $\chi^2$, of all \oi,
\Ni, \sid, and \sit\ lines fitted simultaneously.
In each $\chi^2$ calculation as
a function of the zero level, the same zero level value was adopted
for the \oi\ and \sid\ lines, and for the \Ni\ and \sit\ lines,
since these pairs were observed
in the same sub-exposures. 
In each case, only the \oi\ and \Ni\ lines contributed
significantly to determining the
zero level values, as the lines of \sid\ and \sit\ are too weak. The
\Ni\ lines are sensitive to the zero level as the strongest line is almost
saturated, and because these lines form a triplet and their measured relative
strengths depend on the zero level. 
For our data normalized to unity, the value and
the error bar of the \Ni\ and \sit\ zero levels is: $0.038\pm0.02$, 
{\it i.e.} consistent with zero level within the accuracy of our measurement.
Since the \oi\ line is saturated, the zero level position is sharply
defined.
The measured value of the zero level, for both
\oi\ and \sid\ lines, is: $0.032\pm0.004$. 
Theses values will be used from now on in our analysis and their impact on
the evaluated parameters will be considered as minor in the forthcoming study
and checked in few specific cases. 

\section{The total \hi\ and \di\ content}

\subsection{The stellar Lyman Alpha profile}

  It is well known that the limiting factor of an accurate \dsh\ ratio
measurement is  generally the measurement of the \hi\ column density itself,
as the \di\ line at \lya\ is not saturated for
N(\hi)$\,\sim\,10^{18}\,$\cm2. In the
present case, N(\hi) can be measured with good accuracy from the damping wings of
the \hi\ \lya\ profile. However, we first need to model the stellar continuum, as
the Lorentzian shape of the intrinsic \lya\ photospheric line may mimic 
interstellar damping wings, and  affect the N(\hi) measurement. We
thus  calculated \lya\ profiles from LTE atmosphere models using the appropriate
stellar parameters. Another LTE profile, as well as NLTE calculations,
both using codes completely different from ours, were kindly provided by
Stefan Dreizler. 
A qualitative difference between these profiles is that LTE
profiles are not as sharp and deep as NLTE profiles.

\begin{figure}[h]
\setlength{\unitlength}{1cm}
\centering
%\begin{picture}(9,9)
%\put(0,0){\makebox(8,10){
\psfig{figure=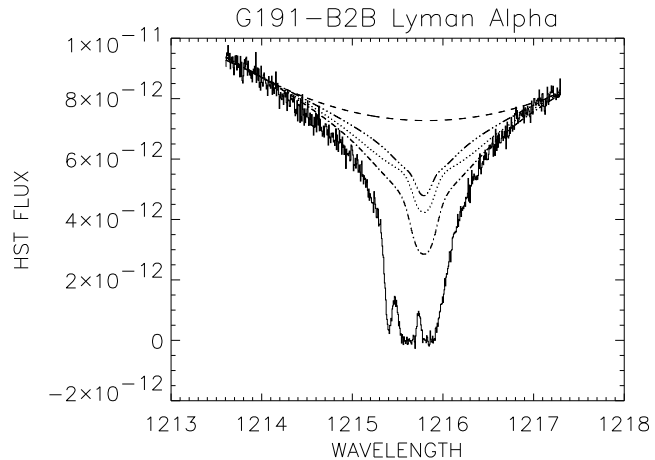,width=\columnwidth}
%}}
%\end{picture}
\caption{The different calculated \lya\ photospheric continua assumed in our 
analysis: from top to bottom, a simple $2^{\rm nd}$ order polynomial, a first LTE
calculation (DK), a second LTE calculation (S. Dreizler,
private communication) and an NLTE calculation (S. Dreizler, private 
communication). These three latter profiles were shifted by 29~\kms\ to match 
the measured radial velocity of G191-B2B.}
\label{lyadatacont}
\end{figure}

 In our analysis, the different possible continuum were
considered (see Fig.~\ref{lyadatacont}) in order to evaluate the possible 
systematic error one could make on the total \hi\ evaluation.

The heliocentric radial velocity of \gbb\ photospheric lines (including
gravitational redshift) is known to be $24\pm5$~\kms\ from ground based 
observations (Reid and Wegner  1988). From our Cycle 1 observations,
we measured a differential velocity $\simeq35\pm4$\kms\ compared to
the nearby interstellar lines (Vidal-Madjar {\it et al.} 1994);
this shift was however evaluated from medium resolution data in which
the ISM components were not resolved. In the present high resolution Echelle-A
data, the photospheric \sit\ line falls at a radial velocity $29\pm1.5$\kms. This
line is however located at the end of our spectral range, and the profile shape
may be partially distorted, inducing a shift in the measured velocity. We also
tentatively identified six Fe~{\sc v} and six Ni~{\sc v} 
photospheric lines in the 1300--1306\AA\ region, whose average location gives a
radial velocity $24.8\pm4.0$\kms. All these measurements are  in good
agreement, and their weighted least square average results in a radial velocity 
(including gravitational redshift): $28.9\pm1.3$\kms. With a gravitational
redshift of order $\sim25$\kms\ (Finley {\it et al.} 1997), the radial velocity of
\gbb\ itself is thus of order $\sim5$\kms\,  a typical value for an object
in the local galactic environment.

  With this information, we produced different data sets for the \lya\ line,
each corresponding to the normalization of the observed data by a theoretical
stellar continuum. From now on, we fix the radial
velocity of the stellar profile at 29\kms. Finally,
we defined an additional ``\lya\ photospheric profile'' with a second
order polynomial, in order to circumvent any assumption made about the
shape of the profile itself. This polynomial continuum is
derived by fitting to the far
wings of the interstellar line in the ranges 1213.6-1213.9 \AA\ and
1217.0-1217.3 \AA\ (see Fig.~\ref{lyadatacont}). 
In fact, we found that the best solution (minimum \ki2) was
obtained for this second order polynomial further corrected by a fourth order
polynomial in the core of the line, the latter being fitted together with the
interstellar \lya\ line. 

  We now have four different profiles, hence four normalized data sets. From
these, we discard the stellar profile corresponding to the LTE calculation of
S.~Dreizler, since, as shown in Fig.~\ref{lyadatacont}, this profile is
intermediate between our LTE calculation, and the NLTE calculation of
S.~Dreizler.
We thus keep these two latter extreme case,
plus the second order polynomial. Hereafter, we respectively denote these
continua as LTE29, NLTE29, and 2$^{\rm nd}$~Polynomial.

\subsection{The total \hi\ and \di\ column densities}

\begin{table}[tp]
\caption[]{Reduced \ki2\ (with 349 degrees of freedom)
and several parameters corresponding to the \lya\ fits 
with the 3 different assumed stellar profiles: $2^{\rm nd}$ order polynomial,
LTE29 and NLTE29 (see text). Only the \hi\ and \di\ lines were fitted, with all
parameters free (see text), and only {\bf one} component on the line of sight.
The order of the polynomial adjusted on top of the reconstructed stellar continuum
is indicated, either 1$^{\rm st}$order or 4$^{\rm th}$order.}
\label{totalHIDI1}
\begin{tabular}
{l@{\excs}l@{\excs}l@{\excs}l@{\excs}l}
\hline
 & & 1$^{\rm st}$order polynomial fit & \\
\hline
 Stellar cont. & $2^{\rm nd}$Pol. & LTE29 & NLTE29 \\
\hline
\ki2 & $1.91$ & $1.51$ & $2.52$ \\
 N(\hi)$_{\rm Tot}$ (\cm2)~ & $3.00\times10^{18}$ &
$2.21\times10^{18}$ & $1.80\times10^{18}$ \\
 N(\di)$_{\rm Tot}$ (\cm2)~& $2.43\times10^{13}$ &
$2.51\times10^{13}$ & $2.66\times10^{13}$ \\
 (\dsh)$_{\rm Tot}$ ($\times10^5$)~ & $0.81$ & $1.14$ & $1.48$ \\
\hline
 & & 4$^{\rm th}$order polynomial fit & \\
\hline
 Stellar cont. & $2^{\rm nd}$Pol. & LTE29 & NLTE29 \\
\hline
\ki2 & $1.13$ & $1.51$ & $2.02$ \\
 N(\hi)$_{\rm Tot}$ (\cm2)~ & $2.32\times10^{18}$ &
$2.21\times10^{18}$ & $2.00\times10^{18}$ \\
 N(\di)$_{\rm Tot}$ (\cm2)~ & $2.60\times10^{13}$ &
$2.66\times10^{13}$ & $2.66\times10^{13}$ \\
 (\dsh)$_{\rm Tot}$ ($\times10^5$)~ & $1.12$ & $1.20$ & $1.33$ \\
\hline
\end{tabular}\\
\end{table}
\begin{table}[tp]
\caption[]{Same as Table~\ref{totalHIDI1} but with {\bf three}
components on the line of sight. All three components are required 
to present a unique \dsh\ value. Here, the total \dsh\ ratio represents the
ratio of the total \di\ column density to the total \hi\ column density.}
\label{totalHIDI3}
\begin{tabular}
{l@{\excs}l@{\excs}l@{\excs}l@{\excs}l}
\hline
 & & 1$^{\rm st}$order polynomial fit & \\
\hline
 Stellar cont. & $2^{\rm nd}$Pol. & LTE29 & NLTE29 \\
\hline
\ki2 & $1.83$ & $1.33$  & $2.05$ \\
 N(\hi)$_{\rm Tot}$ (\cm2)~ & $2.92\times10^{18}$ & $2.12\times10^{18}$ & $1.70\times10^{18}$ \\
 N(\di)$_{\rm Tot}$ (\cm2)~ & $2.58\times10^{13}$ & $2.70\times10^{13}$ & $2.93\times10^{13}$ \\
 (\dsh)$_{\rm Tot}$ ($\times10^5$)~ & $0.88$ & $1.27$ & $1.72$ \\
\hline
 & & 4$^{\rm th}$order polynomial fit & \\
\hline
 Stellar cont. & $2^{\rm nd}$Pol. & LTE29 & NLTE29 \\
\hline
\ki2 & $1.03$ & $1.32$  & $1.71$ \\
 N(\hi)$_{\rm Tot}$ (\cm2)~ & $2.38\times10^{18}$ & $2.01\times10^{18}$ & $1.92\times10^{18}$ \\
 N(\di)$_{\rm Tot}$ (\cm2)~ & $2.73\times10^{13}$ & $2.73\times10^{13}$ & $2.79\times10^{13}$ \\
 (\dsh)$_{\rm Tot}$ ($\times10^5$)~ & $1.15$ & $1.30$ & $1.45$ \\
\hline
\end{tabular}\\
\label{TabHIDI_3}
\end{table}
\begin{figure*}[th]
\setlength{\unitlength}{1cm}
\centering
%\begin{picture}(9,9)
%\put(0,0){\makebox(8,10){
\psfig{figure=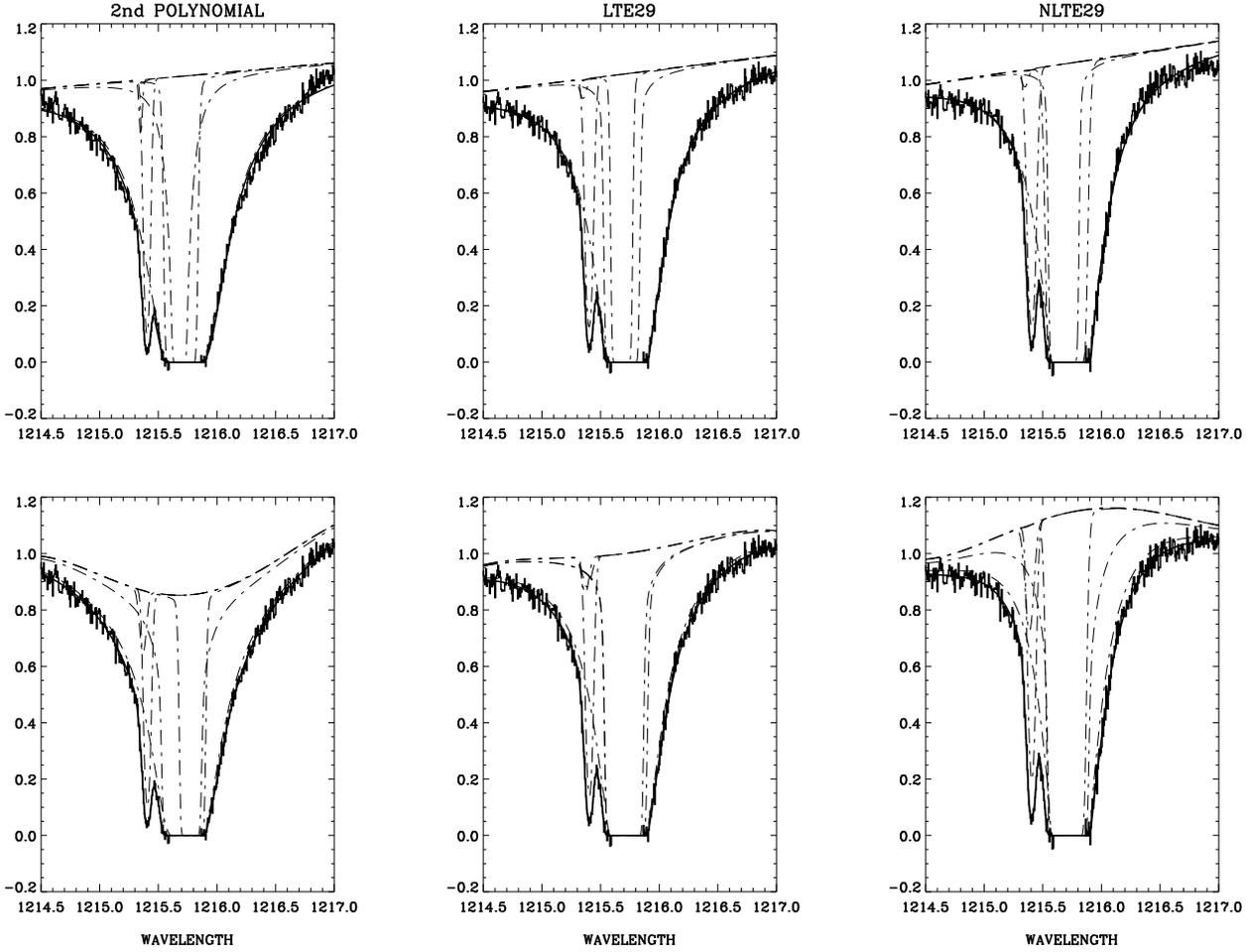}
%}}
%\end{picture}
\caption{The different fits are presented for the \lya\ line when several
stellar continua are assumed. The upper panels correspond to a $1^{st}$ order 
polynomial fit on top of the assumed stellar continuum, while the lower panels
correspond to a $4^{\rm th}$ order polynomial fit.
 From left to right the pannels correspond to the three stellar continua
presented here, {\it i.e.} $2^{\rm nd}$Polynomial, LTE29 and NLTE29 (see text). 
The central panels show that the LTE29 profile is probably a good approximation
to the real stellar photospheric profile, since the 
$4^{\rm th}$ order polynomial fit does not improve the fit with respect to the 
$1^{st}$ order polynomial. On the contrary, the $2^{\rm nd}$Polynomial is  not 
deep enough in the core, while the NLTE29 is  too deep. Once corrected
for their
erroneous shape, the solutions found in terms of total \hi\ and total \dsh\
are  similar (see Table~\ref{TabHIDI_3}). The \ki2\ values  favor
the $2^{\rm nd}\times4^{\rm th}$ stellar continuum (see text).}
\label{lyacont1-4}
\end{figure*}

We now evaluate the total column densities of \hi\ and \di\ along 
the line of sight. We make different assumptions about
the stellar continuum as well as  the number of
components along the line of sight,
in order to evaluate their impact on the estimate
of the column densities. Namely, for each of the above three stellar continua, we
assume a further correction to the continuum, given either by a first order
polynomial, or a fourth order polynomial; in each case, these further corrections
are fitted simultaneously with the physical parameters of the line of sight. As to
the number of components on the line of sight, we performed the calculation for
one, two, three, and four interstellar components. We certainly know from the
other elements that there are at least three absorbing components; what we want to
show here, however, is that the evaluation of the total column densities, and
hence the average \dsh\ ratio, are not
sensitive to the number of components. Finally, in order to further show the
strength of this prediction, we do not take into account any physical constraint
coming from other species, {\it i.e.} \oi, \Ni, \sid, or \sit. When those elements
are included in a global fit of the line of sight, this will narrow down the
margin of freedom for the \dsh\ ratios. 

The results of some of the parameters of these fits are presented in
Table~\ref{totalHIDI1}, for a one cloud solution, and in Table~\ref{totalHIDI3}
for three absorbing components. The results are similar  for a two
component structure of the line of sight. Several comments can be made at this
stage. We note that the
theoretical continua have  very well defined shapes, with abrupt features
in their profile
at various points near the core. When the interstellar absorption is
superimposed, mismatches could be introduced in the resulting final profile.
However, as shown in
Fig.~\ref{lyacont1-4}, the 4$^{\rm th}$ order correction deepens
the 2$^{\rm nd}$ order polynomial in the core of the line to mimic a
Lorentzian profile, but hardly modifies the LTE29 continuum.
This means that the $\Delta\chi^2$,
of order 100, in favor of the $2^{\rm nd}$ order continuum, is mostly intrinsic
to the LTE29 profile.

The above fits also reveal that LTE continua are generally favored
(in terms of \ki2) when compared to NLTE continua.
This is illustrated by the fact that a 4$^{\rm th}$ order correction
plus the NLTE profile tends to the
LTE continuum (see Fig.~\ref{lyacont1-4}).
One could be surprised by this result,
since NLTE model atmospheres should be more appropriate than LTE
calculations at the high temperature of G191-B2B. That
the LTE profiles seem to fit better may partly be due to the fact
that we compare pure hydrogen models only. G191-B2B is well-known
to contain significant amounts of heavy metals, which are visible
in high-resolution UV spectra and which even completely dominate
the EUV spectrum ({\it e.g.}, Wolff {\it et al.} 1998). Taking
into account the blanketing effect of these metals leads to a
decrease of the effective temperature by about 5000 K, and may also
lead to a better agreement of NLTE profiles with the observations.
These details are, however, inconsequential for the present
study.

In summary, these \ki2\ analyses give two main results: the 
$2^{\rm nd}\times4^{\rm th}$ order polynomial is largely favored over the
theoretical continua; it is also supported by the absolute \ki2\ value, with a
reduced \ki2\ close to unity. We can thus conclude that this continuum offers
a very good approximation to the actual stellar continuum of \gbb, and we
adopt this for our determination of the (D/H) ratio. When we include all
other species in the profile fitting, we 
obtain the same result, namely that the $2^{\rm nd}\times4^{\rm th}$ order
polynomial continuum is very largely favored over the other theoretical
continua. Although all subsequent profile fittings were also carried out for
these continua, we will not refer to them any more. Finally,
we also checked that the presence of more than 3 components on the line of
sight does not change these results. Indeed, repeating the same procedure with
4 components led to results that are nearly identical to those of the 3 cloud
solution, in terms of total \hi\ and \di\ column densities.

Given these considerations, we study the relative variation
of the \ki2\ as a function of the total
\dsh\ ratio, in order to derive confidence intervals on this quantity,
according to the so-called $\Delta\chi^2$ method. Strictly speaking, this
method is exact for gaussian quantities, and, more generally, correct up
to second order in the variation of the parameters. The $2\sigma$ confidence
level around a given parameter should then be reached when the \ki2\ of the
fit with the parameter fixed to this $\pm2\sigma$ value becomes equal to 
$\chi^2_0+4$, where $\chi^2_0$ is the total best-fit \ki2\, with the
parameter left as free. The $3\sigma$ level is obtained for $\chi^2_0+9$,
{\it etc..}. However, this assumes that the quantities are gaussian, that
the data points (pixel values) are uncorrelated, and that their error bars
are accurately known. We prefer to relax  these assumptions,
and set the threshold for an ``effective'' $2\sigma$ level at
$\chi^2_0+10$. This provides a {\it very} conservative and {\it very} safe
range of errors. In particular, we note that this procedure has the
great advantage of including in the error bars so-derived all errors on
other parameters that would have to be propagated and projected on the
parameter under study. In effect, when looking for a \ki2\ value for a given
value of the parameter, away from the best fit value, the profile fitting
tends to accomodate the values of the other parameters so as to obtain the
smallest \ki2\, hence the smallest $\Delta\chi^2$.
This is especially important when different parameters are strongly correlated,
{\it e.g.} column densities anti-correlated with broadening parameters, 
or column densities correlated with other column densities.
However, we note that, strictly speaking, this method can only be applied to
statistical errors, and not to systematic errors, which tend to add linearly,
rather than quadratically. Since this method is CPU time expensive,
we only use to determine error bars on \dsh\ ratios.
For other quantities, that do not constitute our primary concern, we will
quote error bars estimates obtained through trial and error.

  Here, from the variations of the \ki2\ as a function of the \dsh\ ratio,
we obtain, for three components, and a unique \dsh\ ratio:

\centerline{(\dsh)$_{\rm Total}$~=~$1.15~\pm0.1~\times10^{-5}$.} 

  In the above, uncertainties arising from the choice of the stellar continuum
are included in the quoted error bars, because of the simultaneous fitted
$4^{\rm th}$ order correction to the continuum. However, systematic errors,
that could arise from uncertain calibration, or physics left out of our model,
are not included.

\section{The component by component analysis}

\subsection{The components velocity separation and column density ratios}

\begin{table*}[tp]
\caption[]{The ISM components characteristics in the direction of \gbb\ 
as  evaluated from the \oi , \Ni , \sid\ and \sit\ lines (from the
blue to the red the components are noted 1, 2 and 3; component 3 is
identified with the LIC and noted as such). The error bars are estimated from
the different fits produced, {\it i.e.} fits with all the lines and 
with all the lines but one included. Note that the error bars on components 1
and 2 are in some cases larger simply because these two components are
blended. As well, note that the values given here will change slightly when
the \hi\ and \di\ lines are included in the profile fitting, see subsequent
tables.}
\label{OINISiIISiIII}
\begin{tabular*}{\textwidth}
{l@{\excs}l@{\excs}l@{\excs}l}
\hline
 & Component 1 & Component 2 & LIC \\
\hline
V (\kms) & $8.2\pm0.8$  & $13.2\pm0.8$  & 
$20.3\pm0.8$  \\
T (K) & $8000.\pm4000.$  & $8000.\pm6000.$  & 
$7000.\pm4000.$  \\
$\sigma$ (\kms)  & $1.5\pm1.5$  & $2.5\pm1.5$  & 
$1.2\pm1.2$  \\
N(\oi ) (\cm2) & $2.8\pm1.0\times10^{14}$  & 
$1.5\pm1.0\times10^{14}$  & 
$3.1\pm0.5\times10^{14}$  \\ 
N(\Ni ) (\cm2) & $1.2\pm0.3\times10^{13}$  & 
$3.5\pm1.0\times10^{12}$  & 
$6.7\pm0.4\times10^{13}$  \\ 
N(\sid ) (\cm2) & $1.9\pm0.3\times10^{13}$  & 
$8.4\pm4.0\times10^{12}$  & 
$1.1\pm0.2\times10^{13}$  \\ 
N(\sit ) (\cm2) & $4.0\pm3.0\times10^{11}$ & 
$2.2\pm0.5\times10^{12}$ & 
$<10^{11}$  \\ 
\hline
\end{tabular*}\\
\end{table*}

 No information on the velocity structure of the
line of sight can be obtained from \lya\, as the components have too large
intrinsic widths in \di\ and \hi\ to be resolved. Therefore, we now turn to a
detailed analysis of the other species, \oi, \Ni, \sid, and \sit,
to extract the required information.

  In order to do so, we perform a fit of all lines of \oi, \Ni, \sid, and \sit,
simultaneously, assuming three components on the line of sight. This means, more
precisely, that we define one velocity, one temperature, and one turbulent
broadening for each cloud; the column densities of the various species are
also free parameters. The absorption profile corresponding to these physical
parameters is then computed for each absorber and for each absorption line, and the
total profile is compared to the observations. We then minimize the sum of the
total \ki2\ of each absorption line. The best fit values of the physical
parameters is presented in Table~\ref{OINISiIISiIII}.
In this table, most of the error bars are tentative, obtained by trial and
error, {\it i.e.} they represent a rough estimate of the actual
error bar; this is however sufficient for our present purpose.

  Among these physical parameters, the radial velocities and the column densities
are the best determined, as would be  expected. The temperature and
turbulent broadening are indeed slightly degenerate as the mass separation between
these species is insufficient. We note that the radial velocities are determined
to a high degree of confidence in the spectral regions where they are resolved,
{\it e.g.} \sit\ for component 2, and \Ni, \oi, and \sid\ for components 1 and 3,
due to the large number of sampling points on the profile. (We recall that the
components are numbered from 1 to 3 in order of increasing radial velocity). The
error on the estimate of their radial velocity is  dominated, in a given
spectral region, by the calibration accuracy $\pm1.5$ \kms. When averaged over all
spectral regions, the accuracy should thus be of order $\pm0.8$\kms. 
The relative shifts in radial velocity between the various components 
are not affected by
the absolute calibration accuracy; their error bars are however 
of order $\pm0.5$\kms but will be greatly improved later.
We obtained these error bars by performing fits to all species but one, each
species being excluded in turn, and averaging the resulting radial velocities
and their shifts.

  It is especially important to note the radial velocity of the third component,
$V=20.3\pm0.8$\kms. This component is identified with the Local Interstellar Cloud
(LIC) in which the Sun is embedded ({\it e.g.}, Lallement \& Bertin 1992; Bertin
{\it et al.} 1995). This cloud has been detected in the direction of nearly all
nearby stars, and its velocity vector relative to the Solar System was found to be:

V$_{\rm LIC} = 25.7$\kms\

l$_{II}$(LIC)$=186.1^o$ 

b$_{II}$(LIC)$=-16.4^o$. 

This component is actually the only absorbing cloud detected toward Capella, whose
line of sight is only separated by $\simeq7^o$ from the \gbb\ line of sight. 
The projections of the LIC velocity on the
Capella and  \gbb\ lines of sight are respectively 21.96\kms, and 20.26\kms. We
therefore obtain a radial velocity in excellent agreement with the prediction.
This value is of importance as it allows us to check the consistency of our
approach: we expect to detect this component, and we know its predicted velocity
within the calibration internal error $\pm0.8$\kms. Taking into account the
appearance of the LIC in four spectral domains (\di, \oi, \Ni, and \sid), we
conclude that we actually well detected and correctly identified it with
component 3. From now on, we thus 
refer to this component as the LIC.

  Finally, we note that the column density ratios between the components are
relatively close to 1. This is illustrated in the following Table 
(from the
blue to the red the components are noted 1, 2 and LIC; component 1 is used as
the reference): 

\bigskip
\begin{tabular}{lll}
\label{compratio}
Components~N~ratios & ~~~2/1~~~ & ~LIC/1~~~\\
\hline
 \oi  & ~~~0.5~~~ & ~~~1.1~~~\\
 \Ni  & ~~~0.3~~~ & ~~~5.6~~~\\
 \sid  & ~~~0.4~~~ & ~~~0.6~~~\\
 \sit  & ~~~5.5~~~ & ~~$<0.02$~~~\\
\end{tabular}\\
\bigskip

This also provides another consistency check. Namely,
when a solution is obtained for \hi\ and \di, the column density ratios between
the different components should remain close to the values obtained for \oi,
and/or \Ni. As far as \Ni, \sid, and \sit\ are concerned, the error on these
ratios, and in particular the LIC to 1 ratio, 
is expected to be relatively small, $\la40$\%, as the lines are not
saturated. Notably, this implies that the various components have very different
ionization structures. Obviously, components 1 and 2 are more ionized than the
LIC. Interestingly enough, \oi\ and \Ni\ do not appear to behave similarly; this
trend will be confirmed below, where a more detailed analysis is performed
including \hi\ and \di.

\subsection{The \hi\ and \di\ content of each component}

\begin{table*}[tp]
\caption[]{Reduced \ki2 , $\Delta$\ki2\ (total) and several parameters corresponding 
to the fits made with the different constraints assumed. All observed lines 
are included in the profile fitting, and all parameters are free, except otherwise
noted in boldface. The first column gives the best-fit all parameters free result;
the subsequent columns give results for various constraints: respectively, for a
unique \dsh\ ratio between the three components, for a unique N(\oi)/N(\hi) ratio,
for a unique N(\Ni)/N(\hi) ratio, for a \dsh\ ratio in the LIC corresponding to
that found by L93, L95, and, finally, for the LIC \dsh\ ratio, temperature and
turbulent broadening corresponding to those found by L93, L95, toward Capella. In
all cases, the stellar continuum is a $2^{\rm nd}\times4^{\rm th}$ order
polynomial.}
\label{soltout}
\begin{tabular*}{\textwidth}
{l@{\excs}l@{\excs}l@{\excs}l@{\excs}l@{\excs}l@{\excs}l@{\excs}l}
\hline
 Constraint & free~~~~~~~~~ & \hi\ follows \di\ & \hi\ follows \oi\ & 
\hi\ follows \Ni\ & (\dsh)$_{\rm LIC}$ & (\dsh ,T,$\sigma )_{\rm LIC}$  \\
\hline
\ki2 $_{\rm Tot}$ & 1.12 & 1.12 & 1.13 & 1.17 & 1.12 & 1.15 \\
$\Delta$\ki2 $_{\rm Tot}$ & 0.0  & 3.9 & 14.6 & 41.8 & 5.8 & 25.6 \\
\ki2 \lya  & 1.04  & 1.05 & 1.06 & 1.14 & 1.05 & 1.07 \\
 N(\hi)$_{\rm Tot}(\times10^{-18})$ (\cm2) & 2.39  & 2.42 & 2.33 & 2.09 & 2.43 & 2.37 \\
 N(\di)$_{\rm Tot}(\times10^{-13})$ (\cm2) & 2.68  & 2.63 & 2.68 & 2.99 & 2.69 & 2.71 \\
 (\dsh$_{\rm Tot}$)$(\times10^5)$ & 1.12 & 1.09 & 1.15 &  1.43 & 1.11 & 1.14 \\
 V$_{1}$ (\kms ) & 8.18 & 8.21 & 8.12 & 8.27 & 8.30 & 8.25 \\
 T$_{1}$ (K) & 11160. & 11027. & 10040. &  12573. & 11213. & 11453. \\
 $\sigma _{1}$ (\kms ) & 0.25 & 0.63 & 1.16 & 0.10 & 0.63 & 0.18 \\
 N(\hi)$_{1}(\times10^{-18})$ (\cm2) & 0.310  & 0.277 & 0.728 & 0.325 & 0.220 & 0.210 \\
 N(\di)$_{1}(\times10^{-13})$ (\cm2) & 0.535  & 0.301 & 0.477 & 0.658 & 0.523 & 0.477 \\
 (\dsh)$_{1}$ ($\times10^5$) & 1.73  & 1.09 & 0.66 &  2.02 & 2.38 & 2.27 \\
 V$_{2}$ (\kms ) & 13.21 & 13.28 & 13.28 & 13.30 & 13.44 & 13.36 \\
 T$_{2}$ (K) & 2575. & 5150. & 3200. & 2650. & 2925. & 3475. \\
 $\sigma _{2}$ (\kms ) & 3.12 & 2.76 & 3.04 & 3.08 & 2.84 & 2.92 \\
 N(\hi)$_{2}(\times10^{-18})$ (\cm2) & 1.08  & 1.27 & 0.430 & 0.103 & 1.52 & 1.42 \\
 N(\di)$_{2}(\times10^{-13})$ (\cm2) & 1.05  & 1.38 & 1.10 & 0.912 & 1.07 & 1.05 \\
 (\dsh)$_{2}$ ($\times10^5$) & 0.97 & 1.09 & 2.56 & 8.85 & 0.70 & 0.74 \\
 V$_{\rm LIC}$ (\kms ) & 20.36 & 20.35 & 20.36 & 20.32 & 20.34 & 20.30 \\
 T$_{\rm LIC}$ (K) & 4160. & 3627. & 5707. & 7467. & 4373. & {\bf 7000.} \\
 $\sigma _{\rm LIC}$ (\kms ) & 2.01 & 2.19 & 1.26 & 0.80 & 1.96 & {\bf 1.60} \\
 N(\hi)$_{\rm LIC}(\times10^{-18})$ (\cm2) & 1.00 & 0.870 & 1.17 & 1.66 & 0.690 & 0.740 \\
 N(\di)$_{\rm LIC}(\times10^{-13})$ (\cm2) & 1.10 & 0.940 & 1.10 & 1.42 & 1.10 & 1.18 \\
 (\dsh)$_{\rm LIC}$ ($\times10^5$) & 1.10  & 1.09 & 0.94 & 0.86 & {\bf 1.60} & {\bf 1.60} \\
 $\Delta V_{2-1}$-$\Delta V_{3-1}$ (\kms ) & 5.03-12.18 & 5.07-12.14 & 5.16-12.24 & 5.03-12.05 & 5.14-12.04 & 5.11-12.05 \\
 \hi\ ratio 2/1-3/1 & 3.48-3.22 & {\bf 4.58-3.14} & {\bf 0.59-1.61} & {\bf 0.32-5.11} & 6.91-3.14 & 6.76-3.52 \\
 \di\ ratio 2/1-3/1 & 1.96-2.06 & {\bf 4.58-3.14} & 2.31-2.31 & 1.39-2.16 & 2.05-2.10 & 2.20-2.47 \\
 \oi\ ratio 2/1-3/1 & 0.57-1.12 & 0.56-1.10 & {\bf 0.59-1.61} & 0.56-1.26 & 0.49-1.10 & 0.52-0.91 \\
 \Ni\ ratio 2/1-3/1 & 0.41-5.61 & 0.40-5.61 & 0.46-5.84 & {\bf 0.32-5.11} & 0.34-5.37 & 0.31-5.13 \\
 \sid\ ratio 2/1-3/1 & 0.44-0.52 & 0.41-0.51 & 0.45-0.50 & 0.41-0.50 & 0.37-0.50 & 0.40-0.51 \\
 \sit\ ratio 2/1 & 5.73 & 4.99 & 5.35 & 4.99 & 4.25 & 4.77 \\
\hline
\end{tabular*}\\
\end{table*}

  We now turn to the final analysis, {\it i.e.} all observed lines fitted
simultaneously. We recall that our fitting procedure works in terms of absorbing
components, so that a unique set of radial velocity, temperature and turbulent
broadening is used to calculate all absorption profiles of all lines for a given
component. These physical parameters are thus fitted simultaneously over the lines
of \di, \hi, \oi, \Ni, \sid, and \sit; we assume the presence of three components
on the line of sight. The total number of degrees of freedom available
is  $\sim800$.

\begin{figure}[h]
\setlength{\unitlength}{1cm}
\centering
%\begin{picture}(9,9)
%\put(0,0){\makebox(8,10){
\psfig{figure=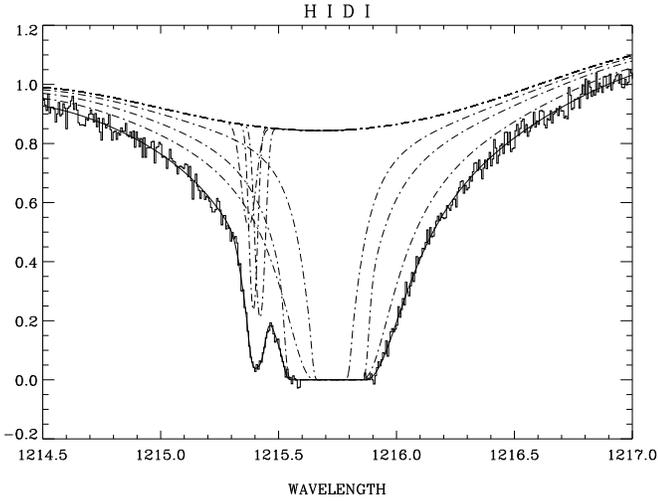,width=\columnwidth}
%}}
%\end{picture}
\caption{The final fit for the \hi\ and \di\ lines when all lines are 
taken into account and all parameters free.}
\label{fitallHIDI}
\end{figure}

\begin{figure}[h]
\setlength{\unitlength}{1cm}
\centering
%\begin{picture}(9,9)
%\put(0,0){\makebox(8,10){
\psfig{figure=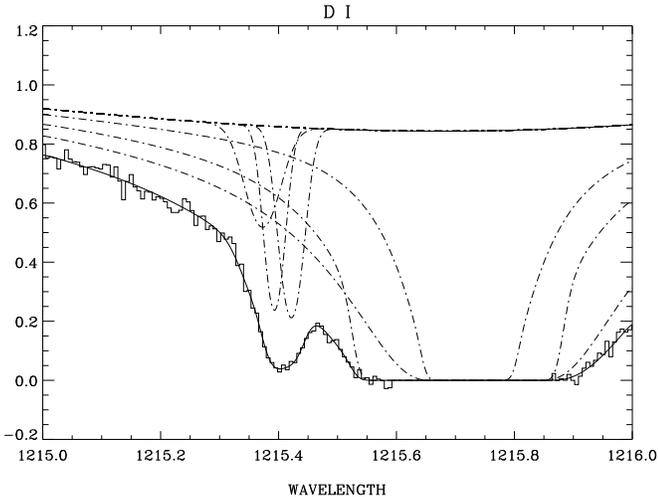,width=\columnwidth}
%}}
%\end{picture}
\caption{The final fit enlarged over the \di\ line when all lines are 
taken into account and all parameters free.}
\label{fitallDI}
\end{figure}

  The best fit, without including any extra constraint on the parameter space, is
shown for the \lya\ line in Fig.~\ref{fitallHIDI}, and is enlarged around the
\di\ line in Fig.~\ref{fitallDI}. This best fit is obtained for a 
2$^{\rm nd}\times4^{\rm th}$ order continuum, and the reduced \ki2 is
$\chi^2=1.12$ for 795 degrees of freedom.  We note that the 
physical parameters obtained, given in Table~\ref{soltout}, 
are in good agreement with those obtained from the fits of the \oi, \Ni, 
\sid, and \sit\ lines only. Moreover, the individual \ki2\ corresponding to
each spectral region, derived here from a simultaneous fit of all lines, are close
to those obtained for individual fits of each spectral region.
This supports the
idea that the structure of the line of sight is well determined.
Finally, we find
column density ratios between components, in \hi\ and in \di, that lie 
relatively close to
those obtained for \oi\ and \Ni\ in the previous fit of \oi, \Ni, \sid, and \sit,
as well as in this global fit. That our results indicate that \hi, \di, \oi,
and \Ni\ behave  in a similar way, represents a strong consistency
check. 
Indeed, it is well known that these four elements have very similar ionization
properties, and, in particular, that they are locked by charge exchange in a
typical ISM environment. Provided that these elements have a unique abundance on
the line of sight, one should therefore expect the abundance ratios of their
neutrals to be unique. 

  We now go one step further with this idea,
and test whether these elements can
actually afford a unique neutral abundance on the line of sight. We thus
require, in turn, the ratios N(\di)/N(\hi), N(\oi)/N(\hi), and N(\Ni)/N(\hi) to be
unique ({\it i.e.}, equal from component to component), and the unique
abundance kept as a free parameter, all other parameters as
equally free as above, and we compare the \ki2\ of
these three solutions with the above best-fit \ki2. 
The results are given in
Table~\ref{soltout}. Obviously, \di\ can be considered as a perfect tracer of \hi\
on this line of sight, as the relative \ki2\ difference with the above best
fit is reasonably small, $\Delta\chi^2=3.9$. Two remarks are in order here. This
result means that, within the quality of our data set, we do not detect variations
of the \dsh\ ratio, from cloud to cloud; it does not mean either that we rule
out such variations. Furthermore, that \di\ seems to trace \hi\ holds here because
we only consider one line of sight. When we compare the value for the \dsh\ ratio
measured here toward \gbb, with that measured toward Capella, in the next section,
we will find that they disagree, and thus argue that the \dsh\ ratio varies within
the local ISM.

 For \oi, the situation is not as clear, as the \ki2\
difference is larger, $\Delta\chi^2=14.6$. However, we note that in this profile
fitting, the instrumental zero-level of the \oi\ line was kept fixed (for numerical reasons).
The \oi\ line is saturated, and the column density ratios between components in
\oi\ thus depend on the value of the zero-level. If one were to incorporate this
additional freedom on the value of the zero-level (within its error bars), the
relative \ki2\ difference would decrease. Therefore we do not feel that the 
above $\Delta\chi^2$ rules out \oi\ as a tracer of \hi. In the case of \Ni,
however, our results indicate that it cannot be considered as a perfect tracer of
\hi, as the \ki2\ difference is quite large, $\Delta\chi^2=41.8$. Varying the
zero-level for the \Ni\ lines cannot account for this large $\Delta\chi^2$, and we
cannot find any other explanation than an actual difference in behavior between
\Ni\ and the other neutrals \di, \oi, and \hi.

  Finally, we wish to point out another important consistency check: whether we
require \di\ or \oi\ to be a tracer of \hi, we obtain the same average
\dsh\ ratio. This supports the robustness of our measurement of the \dsh\
ratio.

\begin{table*}[tp]
\caption[]{Physical characteristics of the ISM components, numbered from 1 to 3 in
the order of increasing radial velocity; component 3 is identified as the LIC, and
noted as such. Here the accuracy on the velocities are 
relative. Absolute accuracy on the heliocentric velocities are of the order of
0.8\kms . Error bars are
tentative 1$\sigma$ estimates, obtained by trial and error.}
\label{comp}
\begin{tabular*}{\textwidth}
{l@{\excs}l@{\excs}l@{\excs}l@{\excs}l}
\hline
Component & 1 & 2 & LIC\\
\hline
V (\kms) & $8.20\pm0.20$  & $13.20\pm0.20$  & $20.35\pm0.05$  \\
T (K) & $11000.\pm2000.$  & $3000.\pm2000.$  & $4000.^{+2000.}_{-1500.}$  \\
$\sigma$ (\kms) & $0.5\pm0.5$  & $3.0\pm0.5$  & $2.0^{+0.5.}_{-1.0.}$  \\
\hline
\end{tabular*}\\
\end{table*}

\begin{figure}[h]
\setlength{\unitlength}{1cm}
\centering
%\begin{picture}(9,9)
%\put(0,0){\makebox(8,10){
\psfig{figure=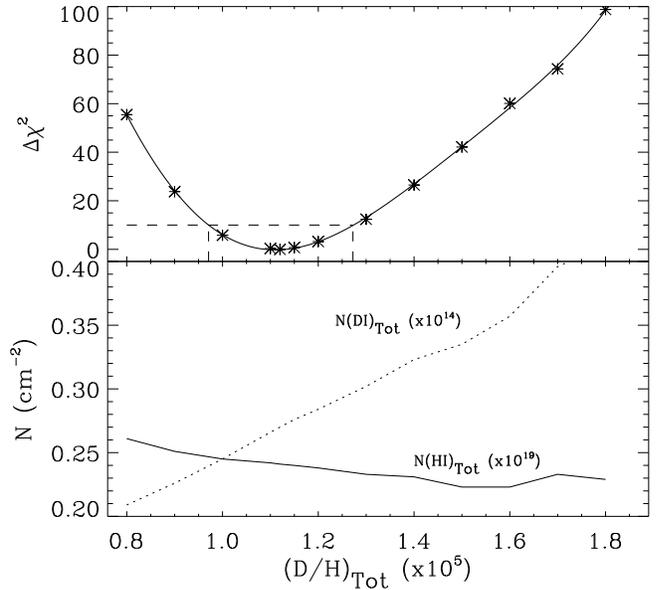,width=\columnwidth}
%}}
%\end{picture}
\caption{Upper pannel: relative \ki2\ difference with respect to the best fit
plotted {\it vs.} the \dsh\ ratio. A unique \dsh\
ratio was assumed on the line of sight. A spline was fitted to the numerical
results, represented by stars. The dashed lines give the minimum and
maximum value of the \dsh\ ratio that correspond to $\Delta\chi^2=10$, or,
according to our conventions, to an effective $2\sigma$ confidence interval. 
Lower pannel: variation of the total \hi\ and \di\ column densities
{\it vs.} the \dsh\ ratio, as indicated.}
\label{Figki2DsH}
\end{figure}

  We derived the  error bar on the \dsh\ ratio by studying the relative
variation of the \ki2\ as a function of the (unique) \dsh\ ratio, as in section
3.2. This variation is shown in Fig.~\ref{Figki2DsH}, where we also plot the
$2\sigma$ error bars estimates corresponding to a $\Delta\chi^2=10$ in the upper
plot, together with the variation of the total \hi\ and \di\ column densities in
the lower pannel. 
We finally obtain the following column densities and \dsh\ ratio with
an ``effective'' $1\sigma$ \dsh\ error bar:

\centerline{N(\hi )$_{\rm Total} = 2.4\pm0.1\times10^{18}$\cm2}
\centerline{N(\di )$_{\rm Total} = 2.68\pm0.05\times10^{13}$\cm2}
\centerline{(\dsh )$_{\rm Total} = 1.12\pm0.08\times10^{-5}$,}

(the error bars for N(\hi) and N(\di) are derived by trial and error; they are
thus more tentative estimates but consistent with the \dsh\ error bar).

\begin{figure}[h]
\setlength{\unitlength}{1cm}
\centering
%\begin{picture}(9,9)
%\put(0,0){\makebox(8,10){
\psfig{figure=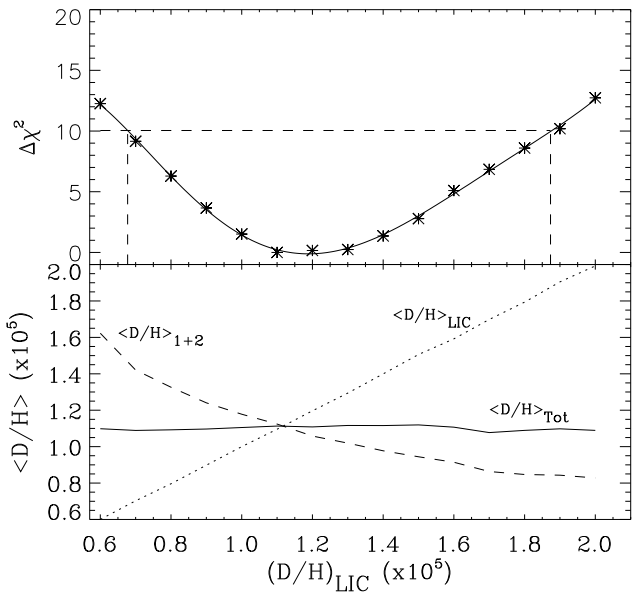,width=\columnwidth}
%}}
%\end{picture}
\caption{Same as Fig.~\ref{Figki2DsH}, except the \dsh\ ratio varied is the LIC
\dsh. In the lower pannel are plotted, as indicated, the evolution of the LIC
\dsh, the \dsh\ averaged over components 1 and 2, and the total \dsh\ averaged
over components 1, 2, and LIC.}
\label{Figki2DsHLIC}
\end{figure}

  Let us now try two further hypotheses on this global line fitting. As we
discussed in the previous section, component 3 is identified as the LIC,
{\it i.e.} the interstellar cloud in which the Solar System is embedded. 
This component is also the unique component detected by L93, L95, 
toward Capella. Clearly, our
average \dsh\ ratio disagrees with their estimate of the \dsh\ ratio toward
Capella, in the LIC: (\dsh)$=1.6\pm0.1\times10^{-5}$.
We now wish to see whether our \dsh\ ratio in component 3, the LIC, 
can be reconciled with
this latter value, and thus proceed to profile fitting, using
(\dsh)$=1.6\times10^{-5}$ in the LIC component. As shown in Table~\ref{soltout},
this solution is indeed compatible with our data set, as the relative 
\ki2\ difference is reasonable, $\Delta\chi^2=5.8$. Obviously, such a high \dsh\
ratio is permitted in the LIC because all three components on the line of sight
to G191-B2B are thoroughly blended in the \di\ line. It also means that estimates
of the \dsh\ ratio in a particular component in our dataset is meaningless, as the
\dsh\ ratios between the three components are strongly correlated. This effect is
shown in Figure ~\ref{Figki2DsHLIC}, where we plotted the $\Delta\chi^2$, the
\dsh\ ratio averaged over components 1 and 2, and the total
\dsh\, {\it vs.} the LIC \dsh. Indeed, the error bars on the LIC \dsh\ ratio, that
correspond (at 2$\sigma$, according to our conventions) to $\Delta\chi^2=10$, are
rather large.
% (\dsh)$_{\rm LIC}=1.19^{+0.35}_{-0.25}\times10^{-5}$ (given here at
%1$\sigma$). 
However, it is quite interesting to note that the \dsh\ ratio of the
LIC, and that averaged over components 1 and 2, are anti-correlated. This
translates itself in that the total \dsh\ ratio remains constant when the
LIC \dsh\ ratio is varied from $0.6\times10^{-5}$ to $2.0\times10^{-5}$.
We obtain the following column densities in the LIC along with the
corresponding \dsh\ ratio with an ``effective'' $1\sigma$ \dsh\ error bar:

\centerline{N(\hi )$_{\rm LIC} = 1.0\pm0.3\times10^{18}$\cm2}
\centerline{N(\di )$_{\rm LIC} = 1.1\pm0.1\times10^{13}$\cm2}
\centerline{(\dsh)$_{\rm LIC} = 1.19^{+0.35}_{-0.25}\times10^{-5}$}

(again, the error bars for N(\hi) and N(\di) are derived by trial and error; 
they are thus more tentative estimates but consistent with the \dsh\ error bar).

  Finally, if we try to impose all characteristics of the LIC found by L93,
L95, toward Capella, in our solution, we obtain
$\Delta\chi^2$=25.6, which seemingly rules out this hypothesis. 
As we just showed
that the LIC \dsh\ ratio was compatible between the two line of sights, we are
forced to conclude that there must be a significant difference in the line
broadening of the LIC between both lines of sight (see however Section 5).

  Let us therefore conclude on this study, as far as \hi\ and \di\ are concerned.
We find that performing a global fit on the line of sight gives good agreement
with the individual fits performed on each spectral region. The average \dsh\ ratio
is tightly constrained to \dsh=$1.12\pm0.08\times10^{-5}$. We also found that \di\ and \oi\
are very reliable tracers of \hi\, within the quality of our data; this is not the
case for \Ni. Moreover, whether \di\ or \oi\ are set as tracers of \hi,
{\it i.e.} a single column density ratio to \hi\ is required on all three
components, then the same average \dsh\ ratio is obtained. This constitutes a
 strong consistency check of our results. Our average \dsh\ ratio is in strong
disagreement with that obtained by L93, L95 toward Capella. The \dsh\ ratio in the
LIC, which is the interstellar cloud common to both lines of sight, can be
reconciled with the \dsh\ found toward Capella. However, this comes at the expense
of a significantly smaller \dsh\ ratio in our components 1 and 2, of order
$0.9\times10^{-5}$, while the \dsh\ ratio averaged all three components remains the
same, (\dsh)$_{\rm Total}=1.12\times10^{-5}$.

\begin{table*}[tp]
\caption[]{Column densities derived in the best-fit solution, obtained
simultaneously on all lines and all species. The third and fifth column,
respectively noted as ``1+2'', and ``Total'', 
give quantities averaged over components
1 and 2, and averaged over all three components, respectively. These are the most
reliable quantities, as components 1 and 2 are blended in all observed lines.
Error bars are tentative estimates, obtained by trial and error. Abundances are
normalized to $10^6$ hydrogen atoms.
%{\bf revoir les barres d'erreurs, elles paraissent un peu petites...
%je n'ai pas change car je ne sais pas trop comment; sur le total, l'erreur
%sur  HI est de moins que 5\%, c'est donc soit pareil soit mieux...}
}
\label{species}
\begin{tabular*}{\textwidth}
{l@{\excs}l@{\excs}l@{\excs}l@{\excs}l@{\excs}l@{\excs}l}
\hline
 Component  & 1 & 2 & 1+2 & LIC & Total \\
\hline
 N(\oi) (\cm2 ) & $2.6\times10^{14}$ & $1.5\times10^{14}$ & $4.0\times10^{14}$ & $2.9\pm0.3\times10^{14}$ & $6.9\pm0.3\times10^{14}$ \\
 N(\oi)/N(\hi) ($\times10^6$) & $823$ & $135$ & $288$ & $287^{+170}_{-80}$ & $288\pm25$ \\
 N(\Ni) (\cm2 ) & $1.1\times10^{13}$ & $4.6\times10^{12}$ & $1.6\times10^{13}$ & $6.3\pm0.4\times10^{13}$ & $7.9\pm0.2\times10^{13}$ \\
 N(\Ni)/N(\hi) ($\times10^6$) & $36$ & $4$ & $11$ & $63^{+33}_{-18}$ & $33\pm2$ \\
 N(\sid) (\cm2 ) & $1.9\times10^{13}$ & $8.4\times10^{12}$ & $2.7\times10^{13}$ & $1.0\pm0.1\times10^{13}$ & $3.7\pm0.1\times10^{13}$ \\
 N(\sid)/N(\hi) ($\times10^6$) & $62$ & $8$ & $20$ & $10^{+6}_{-3}$ & $16\pm1$ \\
 N(\sit) (\cm2 ) & $3.9\times10^{11}$ & $2.2\times10^{12}$ & $2.6\times10^{12}$ & $<1.0\times10^{11}$ & $2.6\pm0.2\times10^{12}$ \\
 N(\sit)/N(\hi) ($\times10^6$) & $1.3$ & $2.1$ & $1.9$ & $<0.1$ & $1.1\pm0.1$ \\
 N(\sid +\sit) (\cm2 ) & $1.9\times10^{13}$ & $1.1\times10^{13}$ & $3.0\times10^{13}$ & $1.0\pm0.1\times10^{13}$ & $4.0\pm0.1\times10^{13}$ \\
 N(\sid +\sit)/N(\hi) ($\times10^6$) & $63$ & $10$ & $22$ & $10^{+6}_{-3}$ & $17\pm1$ \\
 N(\sit)/N(\sid)  & $0.020$ & $0.267$ & $0.096$ & $<0.001$ & $0.070$ \\
\hline
\end{tabular*}\\
\end{table*}

\subsection{Results for each spectral line}

\begin{figure}[h]
\setlength{\unitlength}{1cm}
\centering
%\begin{picture}(9,9)
%\put(0,0){\makebox(8,10){
\psfig{figure=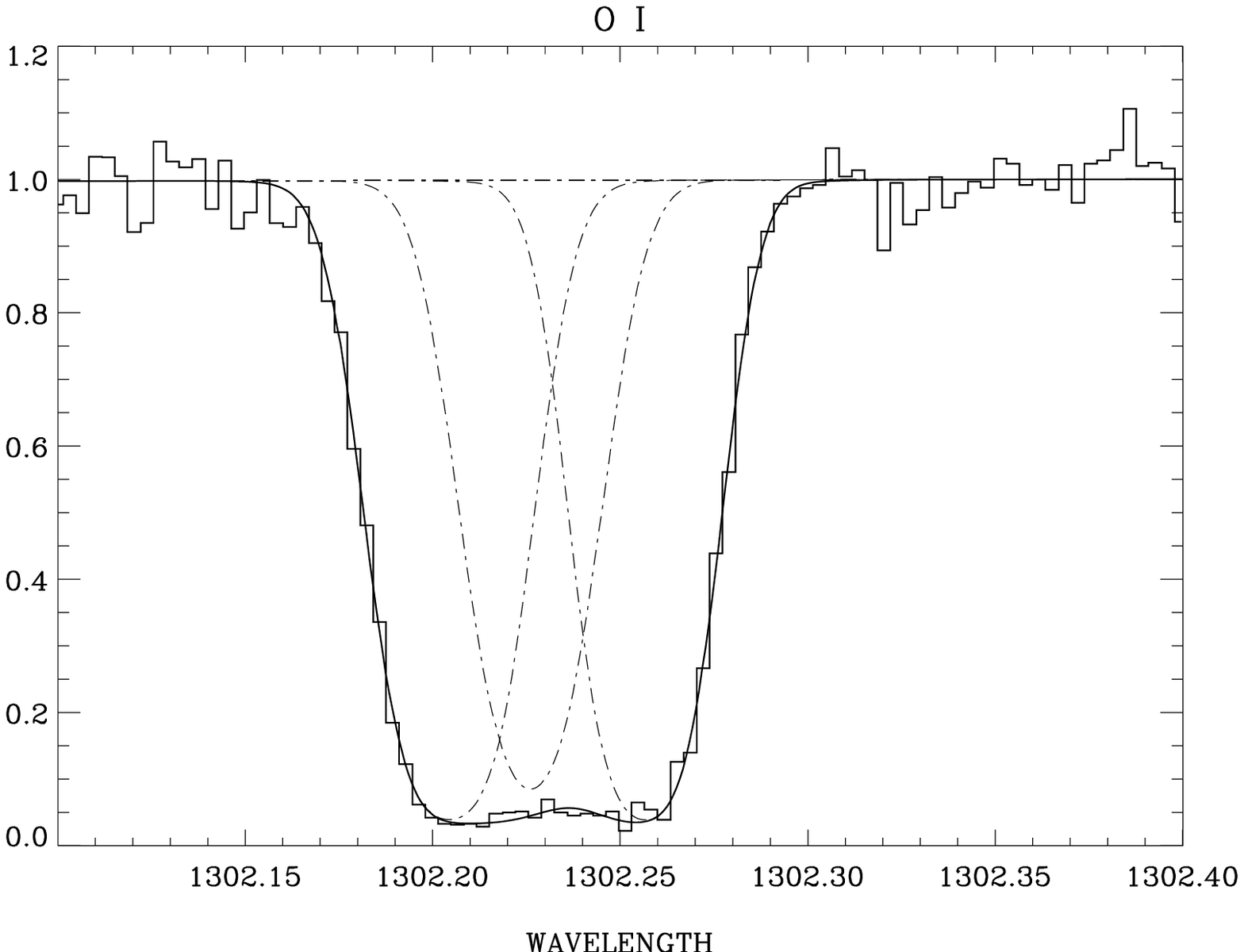,width=\columnwidth}
%}}
%\end{picture}
\caption{The final fit for the \oi\ line when all lines are taken into account
and all parameters free.}
\label{fitallOI}
\end{figure}
\begin{figure}[h]
\setlength{\unitlength}{1cm}
\centering
%\begin{picture}(9,9)
%\put(0,0){\makebox(8,10){
\psfig{figure=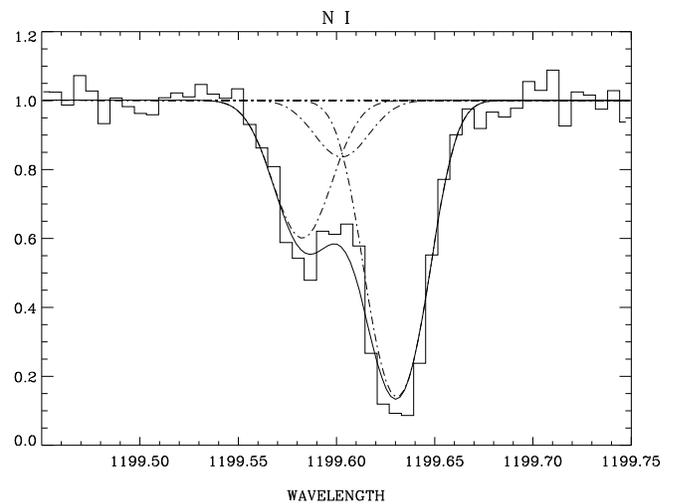,width=\columnwidth}
%}}
%\end{picture}
\caption{The final fit for one of the \Ni\ lines of the triplet 
when all lines are taken into account
and all parameters free.}
\label{fitallNI}
\end{figure}

  We now briefly comment on the final solutions for the lines of \oi, \Ni, \sid, and
\sit, as obtained from the previous global fit of all lines including \hi\ and \di.
These results are summarized in  Table~\ref{species}.

  Components 1 and 2 are blended in all observed lines, and therefore, individual
results for either one of these two components cannot be relied upon. We therefore
discuss quantities averaged over these two components, and refer to them as '1+2'.
The LIC characteristics are obtained with reasonable accuracy,
while total column densities averaged over the line of sight are always the most
accurate.

\bigskip 

\noindent{\it \oi\ 1302\AA:}\medskip

The fit is shown in Fig.~\ref{fitallOI}. The reduced \ki2\ is 0.88
for 144 degrees of freedom, {\it i.e.} the fit is excellent.

The total abundance of \oi\ (normalized to $10^6$ hydrogen atoms)
is N(\oi)/N(\hi)=$288\pm25$, almost identical to that found for the LIC.
This value is also compatible with the average ISM gas neutral oxygen abundance,
given by Meyer {\it et al.} (1998a): N(\oi)/N(\hi)=$319\pm14$.
Meyer {\it et al.} (1998a) also evaluated that the average 
ISM ({\it gas+dust}) should be of the order of $500$ assuming all \oi\
atoms have returned to the gas phase. For comparison the solar value is
N(\oi)/N(\hi)=$741\pm130$ (Grevesse and Noels, 1993), different enough to lead
these authors to conclude that the solar value may be  enriched
in oxygen. From Table~\ref{species}, we see that the LIC appears as a normal
diffuse interstellar cloud, as far as \oi\ is concerned, just as the average of
components 1 and 2. Our observations thus support
the analysis of Meyer {\it et al.} (1998a), with some possible indication
that component 1 has  even returned all its \oi\ to the gas phase.

\medskip

\noindent{\it \Ni\ 1200\AA, triplet:}\medskip

The fit for the strongest line of the triplet is shown in Fig.~\ref{fitallNI}, 
and the total fit over the triplet gives  a reduced \ki2\ per degree of freedom (185) of
1.29. It is certainly less satisfactory
than the \oi\ fit; this seems to result from a slight mismatch between the
three \Ni\ lines. Possibly, this may be due to some blending
with weak photospheric lines of other species 
({\it e.g.}, Fe{\sc v}, Ni{\sc v}, ...). However, if a fit of the triplet
is performed independently of all other lines, the solution is found to be
very similar. We thus remain confident in the robustness of the \Ni\ results.

  Ferlet (1981) found a strong relationship between N(\hi) and N(\Ni) in
a diffuse \hi\ medium, on longer pathlengths, which translates for
$10^6$ hydrogen atoms to N(\Ni)/N(\hi)=$62^{+45}_{-34}$. 
More recently, Meyer {\it et al.} (1998b) reached the following value for the
\Ni\ interstellar abundance: N(\Ni )/N(\hi )=$75\pm4$. These evaluations are
compatible with the solar value of Grevesse \& Noels (1993): 
N(\Ni )/N(\hi )=$93\pm16$. We obtain, as an average over the line of sight, 
N(\Ni )/N(\hi )=$33\pm2$, in agreement with our previous Cycle 1 observations 
(Lemoine {\it et al.} 1996) which gave a value of the order of 32, but
significatively different from both the solar and ISM
values. Interestingly, this difference seems to result from components 1 and 2,
since their average value is $\simeq11$. Furthermore
the LIC nitrogen abundance seems to be compatible with the standard ISM
values: N(\Ni )/N(\hi )=$63^{+33}_{-18}$. As we discuss below, components 1 and 2
appear more ionized than the LIC; this would then suggest that the value of
the 
\Ni/\hi\ ratio is ionization dependent, and could further explain why \Ni\ does not
turn out to be as reliable a tracer of \hi\ as \oi\ and \di\ were found to be.
\bigskip

\noindent{\it \sid\ 1304\AA:}\medskip

\begin{figure}[h]
\setlength{\unitlength}{1cm}
\centering
%\begin{picture}(9,9)
%\put(0,0){\makebox(8,10){
\psfig{figure=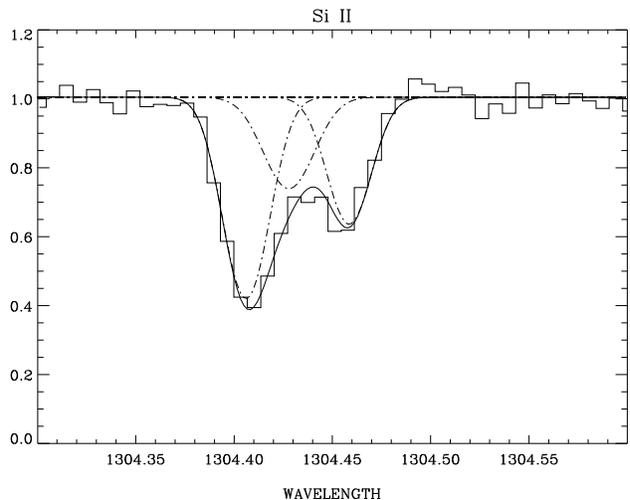,width=\columnwidth}
%}}
%\end{picture}
\caption{The final fit for the \sid\ line when all lines are taken into account
and all parameters free.}
\label{fitallSiII}
\end{figure}

The fit is shown  in Fig.~\ref{fitallSiII}. 
The \ki2\ per degree of freedom 
(72) is here equal to 0.98 and is thus again highly satisfactory.

We find similar column density ratios in \sid, Mg{\sc ii}, and Fe{\sc ii}.
These two latter spectral regions were observed in Cycle 1 
at medium and high resolution (Lemoine {\it et al.} 1996). The correspondence with these
previous observations is as follows: components 1 and 2 are blended
at medium and high
resolution, and correspond to component A in Lemoine {\it et al.} (1996), while,
similarly, component
3 corresponds to component B. The detection of a component C was reported in these
previous observations, but it is not confirmed here. It probably was a ghost due to
the wings of the pre-COSTAR GHRS Echelle-B point spread function. As \sid, 
Fe{\sc ii}, and Mg{\sc ii} have similar ionization properties up to first ionized
level, the similarity of their column density ratios in each component 
certainly supports our results. We discuss below the ionization and depletion of
the three components.
\medskip

\noindent{\it \sit\ 1206\AA:}\medskip

\begin{figure}[h]
\setlength{\unitlength}{1cm}
\centering
%\begin{picture}(9,9)
%\put(0,0){\makebox(8,10){
\psfig{figure=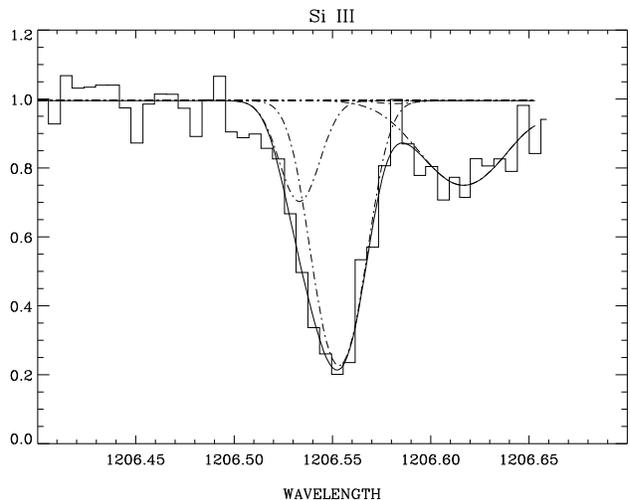,width=\columnwidth}
%}}
%\end{picture}
\caption{The final fit for the \sit\ line when all lines are taken into account
and all parameters free. The broad line on the red side is the photospheric
\sit\ line observed at a 29\kms\ heliocentric velocity. The LIC contribution
to this line is negligible.}
\label{fitallSiIII}
\end{figure}

The fit is shown in Fig.~\ref{fitallSiIII}. The \ki2\ per degree of freedom 
(76) is  1.21 and is thus less satisfactory. However,
this line falls at the end of the spectral range where the S/N ratio in
the continuum varies strongly from pixel to pixel. We did not incorporate this
instrumental effect in our profile fitting procedure. In effect,
we adopted a unique S/N ratio for the whole continuum
(calculated as the average over all pixels in the continuum), and derived
error bars for all pixels using this value of the S/N ratio in the continuum,
weighted by the flux of the pixel, and combined quadratically with the
background noise. Therefore, the variation of the S/N ratio in the continuum,
from pixel to pixel, renders our \ki2\ analysis less meaningful in this case.
In any case, this line contributes to at most 
10\% of the total degrees of freedom (over all lines), so this cannot affect
our global fit.


We confirm the average N(\sit)/N(\sid)$\sim10$\% found in 
Lemoine {\it et al.} (1996). The three components have very 
different N(\sit)/N(\sid) ratios. Components 1 and 2 are more
typical of H{\sc ii} regions, while the LIC is typical of an H{\sc i} region.
The ionization structure of components 1 and 2 is, however, non trivial. 
A N(\sit)/N(\sid)$\sim0.27$ ratio would indicate a 
significantly ionized region, yet the total broadening (thermal plus turbulent)
of the line does not confirm this idea. Indeed, component 1 shows a high
temperature $\sim11000$~K and a low  microturbulence $\sim0.5$~\kms, whereas
component 2 shows the opposite, namely, a low temperature $\sim3000$~K, and a
high microturbulence $\sim3.0$~\kms. These two components may however constitute
differents parts of a shocked structure, with a velocity separation of
$\sim5.0$~\kms\ as projected on the \gbb\ line of sight.
This will be discussed somewhere else.

It is interesting to note that the N(\sit)/N(\sid) ratio of the LIC has
also been evaluated by Gry \& Dupin (1998) in the direction of
$\epsilon$~CMa.
They found N(\sit)/N(\sid)$\sim0.41$, a very different result indeed. If
confirmed, this difference may either show that the ionization structure
within the LIC is inhomogeneous, or that the LIC may be confused by
coincidence with another ISM component on the long $\epsilon$~CMa line of
sight. More observations are needed to clarify this point. 

On the basis of a simple model that calculates the ionization
equilibrium between \sid\ and \sit\ (Dupin and Gry 1998), and assuming 
$n$(\hii)=$n_e$, we converted the \sit/\sid\ ratios listed in Table 7
into hydrogen ionization ratios, N(\hii)/N(\hi), using the most recent
value of n(\hi)$\sim 0.24 \,{\rm cm^{-3}}$ (Puyoo \& BenJaffel 1998).
The derived value are respectively $\sim0.4$, $\sim0.5$,
and $\la 0.1$, for components 1, 2, and the LIC corresponding 
to n$_e$ $\sim0.12$, $\sim0.14$, and $\la0.03$ cm$^{-3}$. Apparently, 
this estimate of the electron density of the LIC is much lower than a
measurement obtained toward Capella using
the ground and excited state lines of C{\sc ii} (Wood \& Linsky 1997):
$n_e=0.11^{+0.12}_{-0.06}$cm$^{-3}$. This discrepancy could be
explained by arguing that the ionization in the LIC is far out of 
equilibrium, due for example to a shock ionizing event, which would 
tend to enhance the current electron density (Lyu \& Bruhweiler 1996).

  We can also use silicon to probe the degree of depletion in the three
components, as it is more sensitive than oxygen or nitrogen. We find a total
nitrogen to neutral hydrogen ratio averaged over the three components of $\sim17$,
{\it i.e.} a factor two smaller than the cosmic value,
$\simeq35$ (per $10^6$ hydrogen atoms). Therefore, provided the ionization of
hydrogen is small, as our calculations seem to indicate, depletion remains
relatively weak. This conclusion can be seemingly applied to the three components
individually, or at least for 1+2 and the LIC, as inspection of
Table~\ref{species} reveals. This low depletion is similar to the one
observed in hot ISM clouds or in the galactic halo (see {\it e.g.} 
Spitzer and Fitzpatrick, 1993, 1995; Fitzpatrick and Spitzer, 1994, 1997;
Jenkins and Wallerstein, 1996; Savage and Sembach, 1996) as well as in the so
called ``CMa tunnel'' (Gry {\it et al.}, 1995; Dupin and Gry, 1998).

  The comparison between the measured radial velocities of the
components and the systemic radial velocity of G191-B2B, $v_{\rm
G191-B2B}=5\pm2$\kms, shows that none of these clouds seems to be associated with
the star. In particular, the association of any of these clouds with an expansion
shell centered on the star is precluded as the clouds have larger radial
velocities than \gbb. A detailed discussion of the  physical 
properties is clearly beyond the scope of the present work. 
This will be the subject of a forthcoming paper.

We~derive a total \hi\ column density of
N(\hi)=2.4$\pm0.1\times10^{18}\,$\cm2 , and confirm the previous evaluation of
Lemoine {\it et al.} (1996). Our value is in disagreement with the EUV
measurements of Kimble {\it et al.} (1993) using the HUT, 
N(\hi)=1.7$\times 10^{18}\,$\cm2 , Green {\it et al.} (1990) 
using a rocket-borne spectrograph,
N(\hi)=1.6$\times 10^{18}\,$\cm2, Holberg {\it et al.}
(1990), N(\hi)=1.0$\times 10^{18}\,$\cm2 , Bruhweiler
\& Kondo (1982) using \Ni\ and the relationship between N(\Ni) and N(\hi), 
N(\hi)=0.8$\times 10^{18}$\,\cm2 . Our estimate is however in agreement within
the error bars with those of Jelinsky {\it et al.} (1988) and
Paerels \& Heise (1989). Moreover, it is in good agreement with the recent EUV
measurements of Dupuis {\it et al.} (1995) and Lanz {\it et al.}
(1996),
N(\hi)=$2.1\times10^{18}\,$\cm2, that use more modern white dwarf atmosphere
models.

The fit of the Lyman limit by Kimble {\it et al.} (1993), Holberg {\it et al.}
(1990) comes through the modeling of the atmosphere of \gbb\ with homogeneous
atmosphere models. Their analysis might thus
be questioned in view of the recent detection of non-negligible
amounts of highly ionized metallic species (N{\sc v}, Fe{\sc v}, Si{\sc iv}) in
this photosphere (see Vidal-Madjar {\it et al.}  1994), although trace of
metallic species are not believed to contribute significantly to the absorption
over 500\AA. We find a total \Ni\ column density in perfect agreement with that 
found by Bruhweiler \& Kondo (1982), and the difference between the N(\hi) values 
is directly related to the difference found in the relationship N(\Ni)--N(\hi)
between our observations and that of Ferlet (1981) on different lines of sight.

\section{Discussion}

  Our results lead us to the following statement: either the \dsh\ ratio is
constant in the local ISM, its value is 1.12$\pm0.08\times10^{-5}$, and the 
estimate made in the direction of Capella by L95 is
incorrect; or, the \dsh\ ratio does vary in the local ISM. If this latter
hypothesis is verified, we need to check whether the \dsh\ ratio can vary
within a same cloud, the LIC, which is common to both the Capella and the \gbb\
line of sight, or whether it varies on larger spatial scales, from
cloud to cloud.

  It thus seems reasonable to re-analyze the evaluation made by L93, L95 
with the same techniques that we used to derive the \dsh\ ratio
toward \gbb. This we do now.

\subsection{Capella revisited}

  The Capella high resolution GHRS Echelle-A data are of high quality, with a
spectral resolution $\Delta\lambda=3.5$\kms, and a signal-to-noise ratio
S/N$\simeq40$ at the interstellar \lya\ line continuum. For this star, the
accuracy on the estimate of the \dsh\ ratio is limited by the ability to
reconstruct the \lya\ stellar emission profile of Capella, all the more since it
is composed of the combined profiles of the two cool stars of the binary system. 
 L95 circumvented this problem by observing Capella
at two different phases, and were thus able to reconstruct, at least partly, both
stellar continua.
%(see Fig.~\ref{Linskycont}). 

  As our paper is not dedicated to the study of this line of sight, we only
summarize our study of the Capella data set, and jump to the conclusions. 
  In order to model the stellar continuum, we used two different approaches, in
the same spirit as Section 3.1: in the first approach, we used the stellar
continuum produced by L95, and in the other, we constructed
a stellar profile in a very naive manner, by interpolating the far wings of the
emission profile with a $7{\rm th}$ degree polynomial. This latter
polynomial would then be
further corrected in the core by a $4^{\rm th}$ order polynomial. In all our
fits, we found that the continuum found by L95 was the most
appropriate in terms of \ki2\ statistics, and hence we will not refer to this
$7^{\rm th}\times4^{\rm th}$ order polynomial continuum anymore. 
L95 detected only one component on the line of sight, from
the analysis of the narrow metal lines of Mg{\sc ii} and Fe{\sc ii}. We followed
these authors and proceeded to the profile fitting with only one component on the
line of sight. A first analysis gave results in agreement with their work,
although we find N(\di)$=2.6\pm0.1\times10^{13}$\cm2, in disagreement with the
original estimate of L95, N(\di)$=2.85\pm0.1\times10^{13}$\cm2. It probably 
results
from a slight mismatch in the radial velocity of the \di\ and \hi\ lines, with a
shift of $0.3$\kms\ for the \di\ line away from the expected position, to be
compared with an error on the radial velocity of only $\pm0.1$\kms. Whereas our
\di\ and \hi\ lines automatically have the same radial velocity, L95 did not link
the radial velocities of the \di\ and of the \hi\ absorbers. Nevertheless, this
seems to point to the existence of another component, that would perturb the \hi\
line but not the \di\ line. 

\begin{figure}[h]
\setlength{\unitlength}{1cm}
\centering
%\begin{picture}(9,9)
%\put(0,0){\makebox(8,10){
\psfig{figure=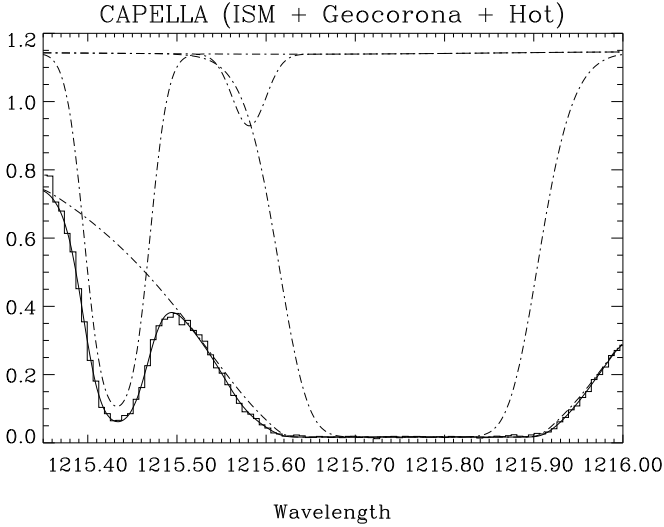,width=\columnwidth}
%}}
%\end{picture}
\caption{The best fit obtained with three components assumed along the
Capella line of sight. The stellar continuum assumed here is the one
reconstructed by L95
and corrected by a 4$^{\rm th}$ order polynomial. The two additional components
are
imposed by the data points near 1215.6\AA\ which cannot be fitted by any
simpler ``one cloud'' solution. The weak additional component needed at
-22\kms\ is at the velocity shift of the earth geocorona but cannot alone fit
correctly the 1215.6\AA\ feature. A third ``hot'' component is also needed
(see text).}
\label{Cap3cloud}
\end{figure}

  Wood \& Linsky (1998) did mention that such a perturbation should be
present, and due to the absorption by the hydrogen atoms in the Earth geocorona,
which, at the time of the observations, was located in velocity space a $-22$\kms.
This absorption is shown in Fig.~\ref{Cap3cloud}. When we attempt a fit
with a two component solution, we obtain a better \ki2\ with a
$\Delta\chi^2\sim100$, a dramatic improvement indeed. In this two component
solution, we left both components entirely free. Interestingly enough, we obtained
two kinds of solution: in both cases, the main component was extremely similar to
the single component (LIC) of L95; in one case, however, the extra absorption was
caused by a cold, {\it i.e.} small broadening component, located around
$\sim-20$\kms; in a second case, slightly favored over the previous one, in terms
of \ki2, we obtained a hot component, with $T\sim30000$K, located around
$\sim+20$\kms. Our profile fitting algorithm indeed works by random
minimization, and can therefore find different degenerate solutions for a same
dataset. Since both of these extra components fall at smaller radial velocities
than the LIC, they produce a shift of the \hi\ absorption, and reconcile the
overall radial velocities of both \hi\ and \di\ lines.

  Since we know that the geocorona has to be present, we performed a three cloud
solution, with all parameters free. We found, in that case, both previous
extra components, {\it i.e.} one hot, located around $\sim+20$\kms, and one cold
component, located at $-22.0$\kms! The final characteristics are given in
Table~\ref{caplya3}, and the fit is shown in Fig.~\ref{Cap3cloud}. The geocorona
found here is entirely consistent with the actual geocorona, in terms of \hi\
column density, velocity, and temperature.

\begin{table}[tp]
\caption[]{Physical parameters evaluated on the line of sight to Capella from
Cycle 1 GHRS archival data. We assumed the presence of three components on the
line of sight, which represents a gain $\Delta\chi^2\sim100$ in $\chi^2$ over the
one cloud solution. The additional components are noted ``Hot'' and ``Geo''
and were left as entirely free in the profile fitting. The data are here
normalized with the stellar continuum reconstructed by L95 and corrected
by a 4$^{\rm th}$order polynomial.}
\label{caplya3}
\begin{tabular}%{\textwidth}
{l@{\excs}l@{\excs}l}
\hline
 Stellar continuum & Linsky $\times4^{\rm th}$ order polynomial \\
\hline
\ki2 /198 & 0.79 \\
 N(\hi)$_{\rm LIC}$ (\cm2 )~~~~~~~~ & $1.72\times10^{18}$ \\
 N(\di)$_{\rm LIC}$ (\cm2 )~~~~~~~~ & $2.69\times10^{13}$ \\
 (\dsh)$_{\rm LIC}$ ($\times10^{5}$) & 1.56 \\
 V$_{\rm LIC}$ (\kms ) & 23.05 \\
 T$_{\rm LIC}$ (K) & 6053. \\
 $\sigma _{\rm LIC}$ (\kms ) & 2.96 \\
 N(\hi)$_{\rm Geo}$ (\cm2 )~~~~~~~~ & $1.74\times10^{12}$ \\
 V$_{\rm Geo}$ (\kms ) & -22.00 \\
 T$_{\rm Geo}$ (K) & 2113. \\
 N(\hi)$_{\rm Hot}$ (\cm2 )~~~~~~~~ & $4.00\times10^{14}$ \\
 V$_{\rm Hot}$ (\kms ) & +21.50 \\
 T$_{\rm Hot}$ (K) & 26667. \\
\hline
\end{tabular}\\
\end{table}

The ``Hot'' component 
falls at an heliocentric velocity of  order  $21.5\pm1.5$\kms, with an
effective broadening temperature $T\sim27000^{+18000}_{-4000}$K. Its
characteristics are well defined, even though it seems to be 
constrained from one side only of the \hi\ line, probably 
because its presence is felt, in these high quality data, over nearly
1\AA. This component may be identified with a cloud interface, as first detected
by Bertin {\it et al.} (1995) toward Sirius~A, or with a ``hydrogen wall'' as
proposed and discussed by Linsky \& Wood (1996), Piskunov {\it et al.}
(1997),  Dring {\it et al.} (1997), and Wood \& Linsky (1998).

Its slightly shifted value redward of the LIC is compatible with the 
``hydrogen wall'' concept since it assumes a slowing down of the ISM hydrogen 
atoms when interacting with the stellar wind itself. If correct, this seems 
to indicate that the LIC extends out to Capella, and that it is 
 quite homogeneous in that direction, since the average volume density 
toward Capella is then very similar to the one
observed within the solar system.

  We expect this component to be entirely compatible with our observations of
\gbb\, as it would fall in between components 2 and 3, in a region where
the \hi\ \lya\ line is completely saturated.

In summary, our estimate of the \dsh\ ratio toward Capella is:

\centerline{(\dsh)$_{\rm Capella} = 1.56\pm0.1\times10^{-5}$,}

in perfect agreement with the original estimate. From these data alone
we also estimated the other characteristics of the LIC independently
and found that its temperature should be T$_{\rm LIC}=6750\pm250$~K
if the microturbulence of the LIC is properly constrained by the other
spectral lines as given in L95.

  As a conclusion, we thus confirm the estimate of L95 of the
\dsh\ ratio within the LIC, (\dsh)$=1.6\pm0.1\times10^{-5}$, but seem 
to evaluate a slightly cooler LIC temperature, more compatible with the
\gbb\ evaluation. The LIC \dsh\ estimate made toward Capella is thus 
marginally compatible with the evaluation made within the LIC in the 
direction of \gbb, and in any case clearly incompatible with the 
estimate of the average \dsh\ ratio on this latter line of sight.
\medskip

{\bf {\it The \dsh\ ratio thus has to vary within the local ISM.}}

\section{Conclusion}

The outcome of our Cycle 5 observations of the white dwarf G191--B2B
performed with the GHRS of HST, aiming to derive "bona fide" value(s) of the 
D abundance in the local interstellar medium is the following : 
Overall 8 absorption complexes of the elements 
\sid, \sit, \Ni, \oi, and, of course, \hi\ and \di, were observed at
high resolution (Echelle-A grating), and analyzed simultaneously using a new
profile fitting procedure. We extract an accurate description of the
cloud structure along the line of sight, which consists of three absorbing
clouds, separated by $\sim$5 \kms. The third absorber, to be understood in
terms of increasing radial velocity, at a heliocentric velocity
$V=20.4\pm0.8$\kms, is identified with the Local Interstellar Cloud in which
the Sun is embedded. The network of constraints on these clouds, that results
from the profile fitting of the different interstellar lines, is such that we
can derive the temperature and the turbulent velocities in the different
interstellar components, as well as the column densities of the various 
observed species.  
Our new set of high quality data is consistent with the Lemoine {\it et al.}
(1996) evaluation of the total \hi\ column density towards
G191-B2B, {\it i.e.} N(\hi)=2.4$\pm0.1 \times10^{18}$cm$^{-2}$.

 We derive  an average \dsh\ ratio over the three absorbing clouds
N(\di)$_{\rm total}$/N(\hi)$_{\rm total}$=1.12$\pm0.08\times 10^{-5}$. 
The Local Interstellar Cloud, detected here toward \gbb, has also been
detected on the line of sight toward Capella, and its \dsh\ ratio
has been measured by Linsky {\it et al.} (1993, 1995):
(\dsh)$_{\rm Capella} = 1.6\pm0.1\times10^{-5}$.
We have  re-analyzed the data of L95 toward
Capella, and confirm their estimate of the \dsh\ ratio, as we find:
(\dsh)$_{\rm Capella} = 1.56\pm0.1\times10^{-5}$.
We find that the \dsh\ ratio in this Local Interstellar Cloud,
on the line of sight to \gbb\, can be made consistent with the Capella value.
However, this comes at the expense of a much smaller average
\dsh\ ratio in components 1 and 2, of order $0.9\times10^{-5}$, in such
a way that the  \dsh\ ratio averaged over all three components remains 
at the above value $1.12\times10^{-5}$. 

  Therefore, we  conclude that the \dsh\
ratio indeed varies in the local ISM over a few parsecs and at least from
interstellar cloud to interstellar cloud. Although we do not detect variations
of the (D/H) ratio within the Local Interstellar Cloud itself, such variations
may become clear as more data are analyzed. Here, we note that both
clouds 1 and 2 are more like \hii\ regions while the Local Interstellar Cloud
could be identified with an \hi\ region; this ionization difference might
point to the  cause of this (D/H) variation.

These large variations of the ISM \dsh\ ratio in so small
scales cannot be due to any nuclear processes (spallation processes induced
by cosmic rays or by huge gamma ray fluxes) occuring in that medium.
We mention here two other possibilities:

(i)- given the fact that the column density is only of the order of a few
$10^{18}$\cm2\ the lowest \dsh\ value could be due to the admixture of a
comparable amount of D-free, pure H remnant of a  material within 
the considered cloud.

(ii)- these low D components could be the remnants of semi-thermalized
blobs of material coming from the outer layers released during the
planetary nebula phase of evolution.

  As a rule, it is much easier to deplete than to create deuterium. In this
case, just as in the two possibilities mentioned above, the lowest ISM
\dsh\ values would be due to a secondary effect, and the highest value
would be more representative of an actual ISM \dsh\ ratio. However, nothing
precludes the Capella value from being itself affected by such a secondary
effect. Moreover, we point out that the deuterium abundance may be locally
enhanced (or depleted) when the radiation flux is anisotropic,
as the radiation pressure
acts differently on \hi\ and \di\ (Vidal-Madjar {\it et al.} 1978; Bruston
{\it et al.} 1981). 

To sum up the consequences of the present investigation, the question 
concerning the existence or not of variations of the \dsh\
ratio in the ISM has now an answer : variations do exist within the local ISM
of order $\approx30\%-50\%$ within a few parsecs,
and/or within a given cloud, the Local Interstellar Cloud.
To be more precise, our observations show that the average (D/H) ratio on
the line of sight to \gbb\ (averaged over three interstellar components),
is significantly different from the (D/H) ratio
measured toward Capella. Although the Local Interstellar Cloud is common to
both lines of sight, we do not find conclusive evidence for variation of the
(D/H) ratio in this particular component.

The comparisons between the theoretical predictions of standard
Big-Bang nucleosynthesis combined with those coming from the recent models of 
chemical evolution of deuterium and the measured \dsh\ ratios in the 
different components of these lines of sight appear to be more complex to
establish. Forthcoming analyses will have now to take into account these
\dsh\ variations and should discuss more thoroughly some of the possibilities
envisaged above.

We therefore look forward to: {\it (i)} further observations of the
interstellar deuterium abundance on other nearby lines of sight. In this
respect, white dwarfs seem to constitute
a highly reliable way of achieving a precise measurement of the deuterium
abundance in the nearby interstellar medium. In effect, white dwarfs
present neither the complex stellar continuum of cool stars,
nor the complex interstellar velocity structure that arises
towards more distant hot stars. {\it (ii)} Measurements of the 
\dsh\ ratio in spiral arms of the Galaxy distinct from ours, or in low-redshift
systems, performed in the framework of the FUSE-LYMAN mission.

\acknowledgements{It is a pleasure to thank S. Burles and D. York for
many stimulating discussions. We also thank J. Linsky and J. Wampler for a
careful reading of the manuscript, L. BenJaffel and O. Puyoo for
ionization calculations, as well as 
S. Dreizler for the calculation of theoretical line profiles. 
We are also endebted to the whole HST Team and in particular to A. Berman 
who has always been extremely helpful in the preparation of the
observations. Work on HST observations in Kiel is supported by grants from
the Deutsches Zentrum f\"ur Luft- und Raumfahrt (DLR).}

\end{document}